\documentclass[letters,usenatbib]{mnras}

\usepackage[T1]{fontenc}

\DeclareRobustCommand{\VAN}[3]{#2}
\let\VANthebibliography\thebibliography
\def\thebibliography{\DeclareRobustCommand{\VAN}[3]{##3}\VANthebibliography}

\usepackage{graphicx}	% Including figure files
\usepackage{amsmath}	% Advanced maths commands
\usepackage{amssymb}	% Extra maths symbols
\usepackage{mathtools}	% multlined environment for multiple lines in equations
\usepackage{xcolor}  % for reviewing purposes

\usepackage{newtxtext,newtxmath}
\title[HI at cosmic noon]{HI content at cosmic noon -- a millimeter-wavelength perspective}

\author[Author]{
Hugo Messias$^{1,2}$\thanks{E-mail: hugo.messias@eso.org},
Andrea Guerrero$^{3}$,
Neil Nagar$^{3}$,
Jack Regueiro$^{4}$,
Violette Impellizzeri$^{5}$,\newauthor
Gustavo Orellana$^{6}$,
Miguel Vioque$^{1,7}$
\\
$^{1}$Joint ALMA Observatory, Alonso de C\'ordova 3107, Vitacura 763-0355, Santiago, Chile\\
$^{2}$European Southern Observatory, Alonso de C\'ordova 3107, Vitacura, Casilla 19001, Santiago de Chile, Chile\\
$^{3}$Departamento de Astronom\'ia, Facultad de Ciencias F\'isicas y Matem\'aticas, Universidad de Concepci\'on, Concepci\'on, 4030000, Chile\\
$^{4}$Princeton University Princeton, NJ 08544-1001 U.S.A.\\
$^{5}$Leiden Observatory, Leiden University, PO Box 9513, 2300 RA, Leiden, The Netherlands\\
$^{6}$Fundaci\'on Chilena de Astronom\'ia, c\'odigo postal 7500011, Santiago, Chile\\
$^{7}$National Radio Astronomy Observatory, 520 Edgemont Road, Charlottesville, VA 22903, USA
}

\date{Accepted XXX. Received YYY; in original form ZZZ}

\pubyear{2023}

\begin{document}
\label{firstpage}
\pagerange{\pageref{firstpage}--\pageref{lastpage}}
\maketitle

\begin{abstract}
%This is a simple template for authors to write new MNRAS papers.
%The abstract should briefly describe the aims, methods, and main results of the paper.
%It should be a single paragraph not more than 250 words (200 words for Letters).
%No references should appear in the abstract.

% context
In order to understand galaxy growth evolution, it is critical to constrain the evolution of its building block: gas. Mostly comprised by Hydrogen in its neutral (HI) and molecular (H$_2$) phases, the latter is the one mostly directly associated to star-formation, while the neutral phase is considered the long-term gas reservoir. 
% aims
%Both phases are difficult to detect directly either due to high excitation temperatures or low transition probability. As a result, while HI direct observations have been limited to the local Universe and extended to high redshifts when seen in absorption, H$_2$ has been traced indirectly via tracers, either Carbon Monoxide (CO) rotational transitions, atomic Carbon fine structure transitions, or dust emission at (sub-)mm wavelengths. However, the latter best tracers the combined content of HI and H$_2$ masses. 
% methods
In this work, we make use of an empirical relation between dust emission at millimeter wavelengths and total gas mass in the inter-stellar medium (M$_{\rm HI}$ plus M$_{\rm H_2}$) in order to retrieve the HI content in galaxies. We assemble an heterogeneous sample of 335 galaxies at $0.01<z<6.4$ detected in both mm-continuum and carbon monoxide (CO), with special focus on a blindly selected sample to retrieve HI cosmological content when the Universe was $\sim2-6\,$Gyr old ($1<z<3$).
% results
We find no significant evolution with redshift of the M$_{\rm HI}$/M$_{\rm H_2}$ ratio, which is about $1-3$ (depending on the relation used to estimate M$_{\rm HI}$). This also shows that M$_{\rm H_2}$-based gas depletion times are underestimated overall by a factor of $2-4$. 
Compared to local Universe HI mass functions, we find that the number density of galaxies with M$_{\rm HI}\gtrsim10^{10.5}\,$M$_\odot$ significantly decreased since 8--12\,Gyr ago.
The specific sample used for this analysis is associated to 20-50\% of the total cosmic HI content as estimated via Damped Lyman-$\alpha$ Absorbers. 
In IR luminous galaxies, HI mass content decreases between $z\sim2.5$ and $z\sim1.5$, while H$_2$ seems to increase.
% conclusions
We also show source detection expectations for SKA surveys.
\end{abstract}


%%%%%%%%%%%%%%%%%%%%%%%%%%%%%%%%%%%%%%%%%%%%%%%%%%

%%%%%%%%%%%%%%%%% BODY OF PAPER %%%%%%%%%%%%%%%%%%

\section{Introduction}
Over the last 25 years it has become increasing clear that the star-formation (SF) history in the Universe peaked at about 10\,Gyr ago \citep{Lilly96,Madau98,HopkinsBeacom06,MadauDickinson14}. Nevertheless, it is only in the last 5 years, with the advent of spectral scan surveys conducted with Atacama Large (sub-)Millimeter Array \citep[ALMA;][]{Brown04}, \textit{Jansky} Very Large Telescope \citep[\textit{JVLA};][]{Perley11}, and the NOrthern Extended Millimeter Array \citep[NOEMA;][]{Guilloteau92}, that the community has identified the driver of this effect in a statistically consistent manner. Namely, the cosmological content of molecular-Hydrogen (H$_2$) in the Inte-Stellar Medium (ISM) peaks at a similar epoch \citep{Aravena2016,Decarli19,Decarli20,Riechers19,Lenkic20}.

After the fact, this may come as no surprise since: it is long known that star-formation rate surface density correlates well with total ISM gas -- in either neutral or molecular phases --- surface density \citep[the so-called Schmidt-Kennicutt, SK, law;][]{Schmidt59, Kennicutt98}; gas in regions with high SF efficiency is mostly in its molecular phase, while in less SF efficient regions gas is mostly in its neutral phase (HI) \citep{Bigiel08}; and the SK law holds up to high redshifts \citep[e.g.,][]{Bouche07}.

Nevertheless, both neutral and molecular Hydrogen components in galaxies are difficult to directly detect with increasing redshifts. The former, because of its low transition probability, while the former due to the absence of a permanent dipole moment. Hence, indirect tracers are required if one aims to estimate their content in galaxies up to early cosmic times.

Historically, H$_2$ has been traced via Carbon Monoxide (CO) emission, with the ground based rotational transition (J:1-0) being the reference tracer. If a higher-J transition is available, population-wide line ratios (R$_{\rm JJ-10}$) are usually adopted to estimate the ground transition emission \citep[e.g.,][]{CarilliWalter13}. The H$_2$ mass can then be obtained from the CO-luminosity by considering a conversion factor ($\alpha_{\rm CO10}$) that remains a topic of discussion. Today, it is agreed that this factor is highly metallicity-dependent, and that in Solar-like metallicity environments, one can adopt the value found in the Milky-Way \citep[see][and references therein]{Dunne22}. However, uncertainties in both R$_{\rm JJ-10}$ and $\alpha_{\rm CO10}$ (among others) can easily build up and result in significant uncertainties in the estimated molecular gas mass (M$_{\rm H_2}$). With the advent of deep observations with ALMA and NOEMA, the community also now considers neutral Carbon (CI) as a molecular gas tracer \citep{GerinPhillips00,Papadopoulos04,Tomassetti14}, making use of its forbidden transitions [CI]${\rm ^3P_1-^3P_0}$ and ${\rm ^3P_2-^3P_1}$. Despite being fainter than CO low-J transitions, its spectral line energy distributions (SLED) is simpler (only three levels), it is optically thin in most extra-galactic environments, and has fewer excitation mechanisms than CO \citep[see][and references therein]{Dunne22}. Nevertheless, this alternative also requires a conversion factor with an associated uncertainty and considerable telescope time.

On the other hand, neutral Hydrogen has only been directly detected up to $z\sim0.2-0.4$ \citep{Lah07,Fernandez16}, and up to $z=1.3$ with the aid of strong gravitational lensing \citep{ChakrabortyRoy23}. Otherwise, at high-redshifts, neutral gas has been mostly traced in absorption \citep{Rao06,Braun12,Zafar13} and trace impact parameters in the 0.1--1\,Mpc range, significantly larger than the typical visible galaxy size regardless of cosmic time or reference wavelength (i.e., galaxies at high-redshift are intrinsically smaller, \citet{Buitrago08}, while the largest known \textbf{\color{teal}spiral} galaxy is about 200\,kpc in diameter \citet{Galaz15}). Alternative approaches comprise methods making use of optical continuum or spectroscopic properties \citep[e.g.,][]{Zhang09,Catinella12,Brinchmann13,Parkash18,Bera22}, or using [CII]\,158\,$\mu$m as a HI tracer \citep{Heintz21}, or combining the SK law with the relation between molecular gas fraction and the mid-plane pressure acting on a galaxy disc \citep[][making use of \cite{Bigiel08,BlitzRosolowsky06}]{Popping15}.

In this work, we present a millimeter-wavelength-based method to estimate the neutral Hydrogen content in galaxies. In Section~\ref{sec:method}, we detail the methodological basis of our approach. In Section~\ref{sec:sample}, we detail the sample assembled from the literature considered in our analysis. In Section~\ref{sec:results}, we report our results including the derived HI cosmological content at $1 \lesssim z \lesssim 3$, while in Section~\ref{sec:disc} we discuss these results. The conclusions are presented in Section~\ref{sec:conc}. The adopted cosmology refers to the results reported by \citet[][\textit{Planck18} cosmology henceforth]{Planck20}, namely ${\rm H_0=67.7\,km ~ s^{-1} ~ Mpc^{-1}}$ and ${\rm \Omega_M=0.31}$ (note that in some figures we also use the Hubble constant dimensionless equivalent $h_{P18}=0.677$).

%% ==== METHODOLOGY
\section{Methodology} \label{sec:method}

\subsection{Estimating ISM gas mass via dust continuum emission} \label{sec:dust2ism}

The last decade has shown the potential of using the mm continuum emission of galaxies to retrieve their molecular-gas content up to very high redshifts \citep[][]{Scoville14,Scoville16,Scoville17,Magnelli20}. This is based on the fact that the Rayleight-Jeans tail of the dust thermal emission in galaxies is mostly optically thin, thus being a tracer of the total dust mass \citep[][and references therein]{Scoville14,Orellana17}. %Hughes17
The latter is related to the star-formation activity \citep[both its production and heating;][]{KennicuttEvans12}, which is observed to be related to the molecular gas content \citep{Schmidt59, Kennicutt98, Bouche07, Bigiel08}. This mm-continuum-based conversion factor (referenced to rest-frame 850~$\mu$m, $\alpha^{\rm H_2}_{\rm 850}$) has the great advantage that it requires significantly less telescope time while providing an estimated molecular-gas mass density evolution in the Universe in very good agreement with those based in line observations (Section~\ref{sec:mfdens}).

The underlying assumption of $\alpha^{\rm H_2}_{\rm 850}$ is that dust is well mixed with the molecular gas and mostly traces H$_2$-dominated regions. However, this scenario is likely only applicable to the luminous-end of the KS-law regime \citep{Bigiel08}, and it has been shown that the M$_{\rm dust}$ (and as a result 850\,$\mu$m luminosity, L$_{850}$, as a proxy) yields a tighter relation with the total gas mass \citep{Orellana17,Casasola20}, thus including neutral gas (HI). In this work, total gas mass in the ISM refers to:
\begin{equation}
    {\rm M_{ISM} = \alpha_{heavy}~(M_{HI}+M_{H_2})}
    \label{eq:mism1}
\end{equation}
where $\alpha_{\rm heavy}=1.36$ is the factor that corrects the gas mass estimate in order to account for chemical elements heavier than Hydrogen in the ISM \citep{Croswell96, CarrollOstlie06}, and M$_{\rm H_2}~=~\alpha^{\rm H_2}_{\rm CO}~L'_{\rm CO10}$. Unless otherwise stated, we adopt $\alpha^{\rm H_2}_{\rm CO}=2.9$, which is the value reported by \citet[][$\alpha^{\rm H_2}_{\rm CO}=4.0$]{Dunne22}, but uncorrected for heavier element fraction.

In order to establish a relation between M$_{\rm ISM}$ and L$_{850}$ we have adopted the sample used in \citet[][and references therein]{Orellana17}. Briefly, it comprises different types of galaxies found in the local-Universe ($z<0.1$) for which reliable HI and CO\,(1-0) detections are available in the literature. The HI fluxes are global ones, as well as the CO measurements obtained with single-dish facilities. The global dust continuum measurements at 850\,$\mu$m come from \textit{Planck} mission and are corrected for: (i) systematic offsets observed when compared with SCUBA-850 observations \citep{Orellana17}; (ii) spectral shape corrections (a multiplicative factor of 0.887 for a spectral-index of 3; Table~14 in the Explanatory Supplement to the
Planck Early Release Compact Source Catalogue\footnote{https://lambda.gsfc.nasa.gov/data/planck/early\_sources/explanatory\_supplement.pdf}); and (iii) galactic CO\,(3-2) contamination \citep[3\%;][]{Orellana17}. For interpretation purposes of the results, it must be emphasized that this ``global-measurement'' approach means the fluxes of the different components are not fully co-spatial, whereby we do include in this analysis HI emission from regions where no dust or CO are detected.

\begin{figure*}
    \centering
    \includegraphics[width=0.49\textwidth]{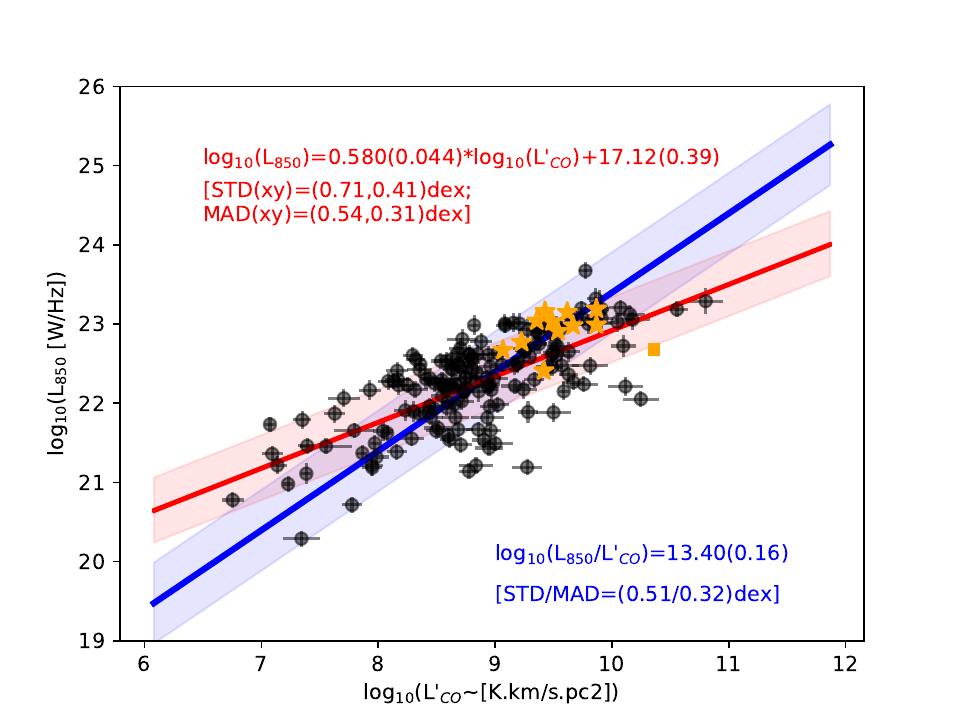}
    \includegraphics[width=0.49\textwidth]{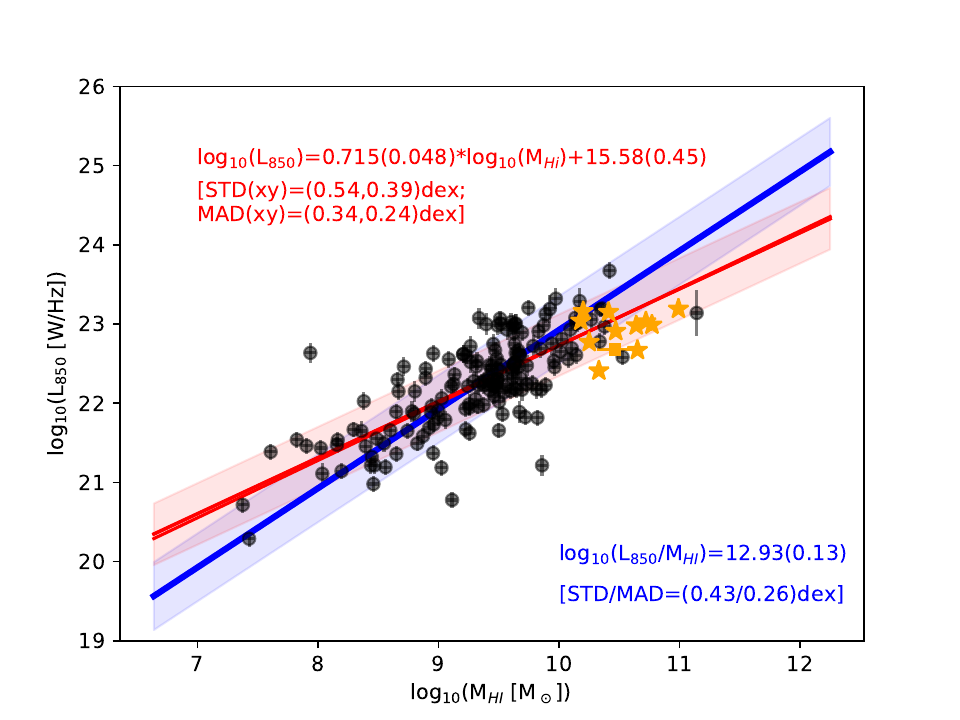}\\
    \includegraphics[width=0.49\textwidth]{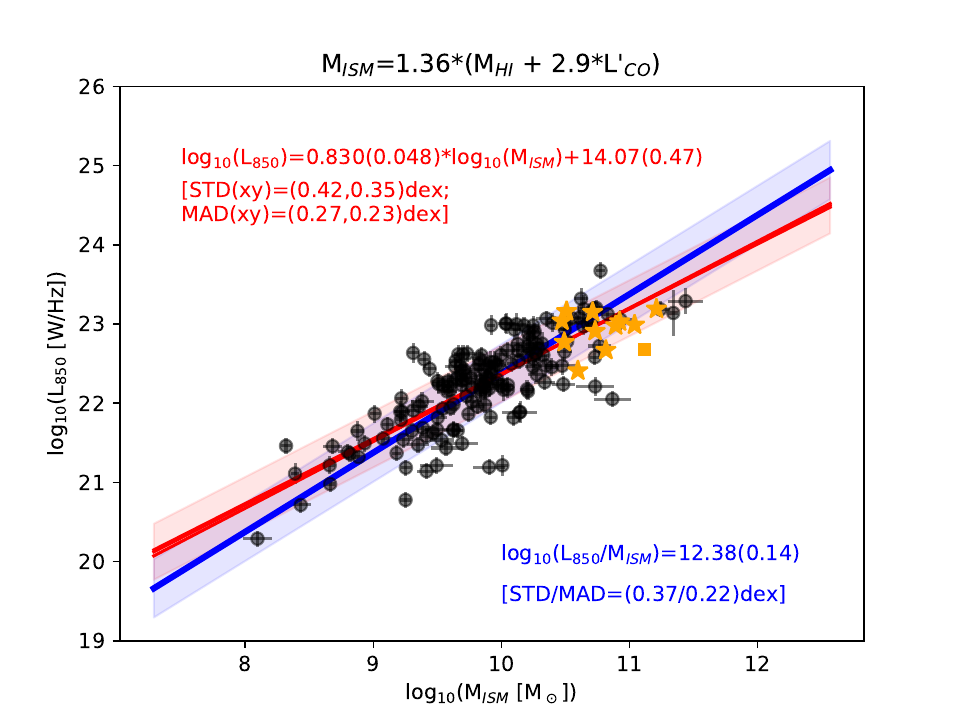}
    \includegraphics[width=0.49\textwidth]{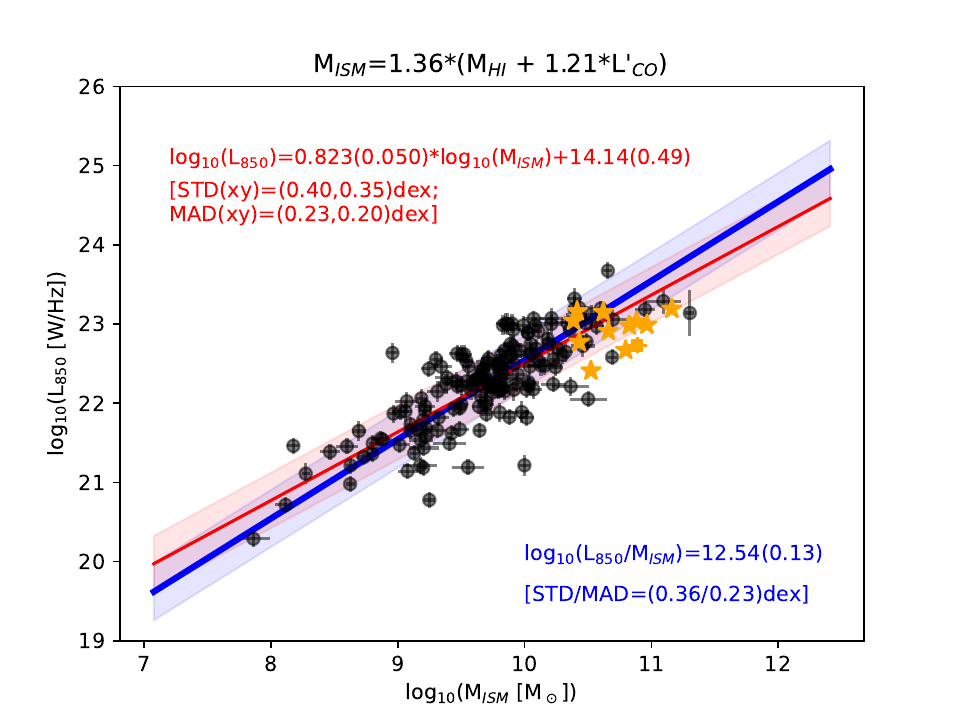}
    \caption{For the sample used in \citet[][black circles]{Orellana17}, we show how L$_{850}$ relates to $L'_{\rm CO10}$ (as a tracer for M$_{\rm H_2}$; top left-hand side panel), M$_{HI}$ (top right), M$_{\rm ISM}$ (bottom left; Equation~\ref{eq:mism}), or M${\rm _{ISM}^R}$ (bottom right; Equation~\ref{eq:mismR}). The blue line and region show the overall parameter ratio and the data standard deviation ({\sc std}) in the log-log space. The red line and region show the best linear fit to the data and its {\sc std} in the log-log space. The results are presented in each panel in the bottom right (ratio) and upper left (linear fit), together with the {\sc std} and the median absolute deviation ({\sc mad}) of the population in the \textit{x} and \textit{y} directions. The results from the bottom-right panel are the ones reported in Equations~\ref{eq:mismRT} and \ref{eq:mismLR}. An higher-redshift sample \citep{Cybulski16, Cortese17, Fernandez16} is shown for reference (yellow data points), and intentionally not used in the analysis.}
    \label{fig:dust2ism}
\end{figure*}

In Figure~\ref{fig:dust2ism} we specifically show how L$_{850}$ relates with ${\rm M_{H_2}}$ (upper left panel), ${\rm M_{HI}}$ (upper right), and ${\rm M_{ISM}}$ (lower left). Two type of fits are shown in the panels, a log-log linear relation ($\log(L_{850}) = m~\log(M_{gas}) + C$; red line and shaded region; referred to as LR method henceforth) and a 1-to-1 ratio ($\log(L_{850})/\log(M_{gas})$; blue line and shaded region; RT method henceforth). %Throughout this manuscript, we will refer to them as LR and RT, respectively. 
The lower-right panel also shows L$_{850}$ versus ${\rm M_{ISM}}$, but where $\alpha^{\rm H_2}_{\rm CO}$ was also left free in the fitting process (note that no priors were used to constrain the value). In each panel, we report the best fits (LR in red, RT in blue, the values in parenthesis shows the fit parameters uncertainties) together with the sample's standard deviation ({\sc std}) and median absolute deviation ({\sc mad}) in the x- and y-axis directions. For reference, we also show higher-redshift galaxies (yellow stars and square) for which there are also direct detections of HI, CO, and dust continuum \citep{Cybulski16, Cortese17, Fernandez16}, but were not used in the derived fits.

Quantitatively, based on each fit's resulting population {\sc std} and {\sc mad} values, it is clear that there is a fitting improvement from left to right, and top to bottom, with overall reductions in {\sc std} and {\sc mad} of 0.3\,dex and 0.1\,dex, respectively, for RT. The differences for the LR results are even larger. In the lower-right panel, the reported best-fit value for $\alpha^{\rm H_2}_{\rm CO}$ is $1.21\pm0.10$. The improvement in the sample spread while using this value is small ($<0.1\,$dex), but we interpret this reduction in $\alpha^{\rm H_2}_{\rm CO}$ as a balancing of the neutral and molecular regions that L$_{850}$ is tracing. In other words, there are galaxies or regions in galaxies where ${\rm M_{H_2}}$ and ${\rm M_{HI}}$ amount to similar quantities/masses. Considering them together would thus mean doubling the weight of such regions when deriving a fit (i.e., lower-left panel). As such, we interpret this value of $\alpha^{\rm H_2}_{\rm CO}~=~1.21$ as the limit identifying the regions in galaxies in which L$_{850}$ is tracing a dominant ${\rm M_{H_2}}$ component of the ISM gas. However, this interpretation needs a more careful analysis, and is deferred to future work. In this manuscript, since we wish to test the implications of using either RT or LT, we adopt $\alpha^{\rm H_2}_{\rm CO}~=~1.21$ in the adopted RT and LT, but $\alpha^{\rm H_2}_{\rm CO}~=~2.9$ when estimating total ${\rm M_{H_2}}$. As a result, we adopt the following nomenclature throughout the manuscript:
\begin{equation}
    {\rm M_{ISM} = \alpha_{heavy}~(M_{HI}+2.9~L'_{CO10})}
    \label{eq:mism}
\end{equation}
\begin{equation}
    {\rm M_{ISM}^R = \alpha_{heavy}~(M_{HI}+1.21~L'_{CO10})}
    \label{eq:mismR}
\end{equation}
and the following relations:
\begin{equation}
    {\rm RT \equiv \log\biggl(\frac{L_{850}[W/Hz]}{M_{ISM}^R[M_\odot]}\biggr) = 12.54(\pm0.13)}
    \label{eq:mismRT}
\end{equation}
\begin{equation}
    {\begin{multlined}[b] \rm LR \equiv \log(L_{850}[W/Hz]) = 0.823(\pm0.050)~ \log(M_{ISM}^R[M_\odot]) \\ + 14.14(\pm0.49) \end{multlined}}
    \label{eq:mismLR}
\end{equation}

\subsection{Rest-frame 850~$\mu$m continuum estimates}

We have considered two different approaches in order to estimate the dust continuum flux density at rest-frame 850~$\mu$m ($S_{850}^{\rm RF}$): (i) modified black body fit to FIR plus (sub-)mm photometry; (ii) power-law fit to (sub-)mm photometry ($S_\nu \propto \nu^\beta$). The former was pursued by making use of {\sc mbb emcee}\footnote{https://github.com/aconley/mbb\_emcee} \citep{ForemanMackey13,Conley16} to fit \emph{Herschel} Space Observatory \citep[\emph{Herschel};][]{Pilbratt10} photometry together with 0.85--3\,mm photometry when available. However, we soon found that this method overestimated $S_{850}^{\rm RF}$ when comparing the fit with and without (sub-)mm photometry. Sometimes, even considering the latter, the \emph{Herschel} photometry weighted more to the fit, resulting in an overestimated $S_{850}^{\rm RF}$. As a result, we do not consider those galaxies for which a (sub-)mm detection is not reported.

\begin{figure}
    \centering
    \includegraphics[width=0.48\textwidth]{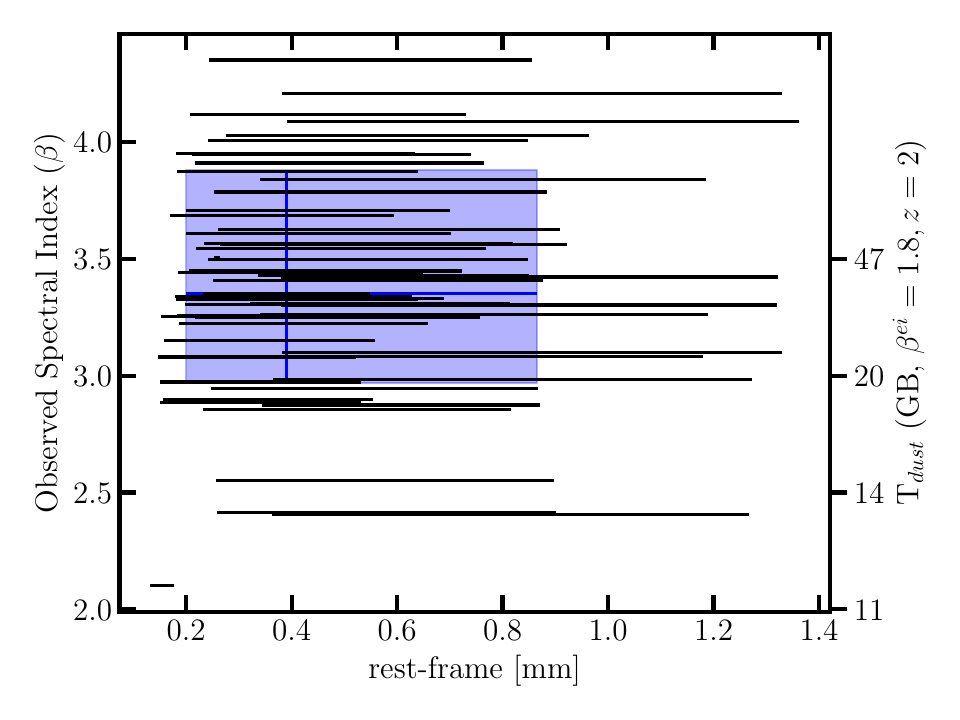}
    \caption{The distribution of the estimated spectral-index in the observed mm spectral range (0.8--3\,mm) for 56 galaxies in our sample, for which there are more than one mm-photometry data points. The solid lines show the rest-frame wavelength range used to estimate the spectral-index. The blue region and errorbar report the median and 16$^{\rm th}$ and 18$^{\rm th}$ percentiles range. The outlier with a low spectral-index is a COLDz galaxy at $z_{\rm spec}=5.3026$, which is the sole case where the photometry is not tracing the RJ regime. For reference, the right-hand side $y$-axis shows the dust temperatures reproducing the spectral indices on the left axis. For this exercise, we have adopted a gray (modified black) body model at $z=2$ with an emissivity index $\beta^{ei}=1.8$, estimating the spectral index between 0.6 and 2.6\,mm (observed frame; values matching the 16$^{\rm th}$ and 18$^{\rm th}$ percentiles range). We have also considered the effect of a hotter CMB to retrieve observed spectral indices (Section~\ref{sec:cmb}).}
    \label{fig:specidx}
\end{figure}

Approach (ii) was thus the only adopted approach to estimate $S_{850}^{\rm RF}$. In Figure~\ref{fig:specidx} we show the spectral-index distribution of the 56 sources (47 are from the \citet{Birkin2021} sample; Section~\ref{sec:b21flux}) for which more than one frequency photometry is reported (the minimum and maximum rest-frame are shown with the solid line). The median value of $\beta$ is 3.35$^{+0.53}_{-0.38}$ ({\sc mad}$=$0.26) shown as the blue region and errorbar in the figure. This was the spectral-index adopted in cases where only one (sub-)mm photometry data point is available. We note that it is common to use $\beta=3.8$, while lower values are associated with estimates based on higher rest-frame frequency values closer to the peak of the dust black-body emission. To test this, we have limited the analysis to the 11 galaxies for which the lower- and upper-wavelength photometry trace rest-frames longer than 300 and 850\,$\mu$m, respectively. The results remain the same within the uncertainties: $\beta=3.26^{+0.68}_{-0.32}$.

We do a final comparison between the adopted spectral-index flux estimate approach and the case when one normalizes a gray (modified black) body (GB) only to the millimeter spectral range flux estimates (i.e., not considering the FIR spectral range as mentioned above). The spectral energy distribution is thus described with \citep{Hildebrand83}:

\begin{equation}
    I_\nu ({\rm T_d},z) \propto \kappa_{\nu_{rest}} B_\nu(T_d,z)
\end{equation}

where $B_\nu$ is the black body function assuming a dust temperature $T_d$, the dust mass absorption coefficient is $\kappa_{\nu_{\rm rest}} \propto \nu_{\rm rest}^{\beta^{ei}}$, and $\beta^{ei}$ is the dust emissivity index. Following \citet{Scoville16}, we adopt $T_d=25\,$K and $\beta^{ei}=1.8$. We further consider the effect of a hotter CMB (Section~\ref{sec:cmb}).

\begin{figure}
    \centering
    \includegraphics[width=0.48\textwidth]{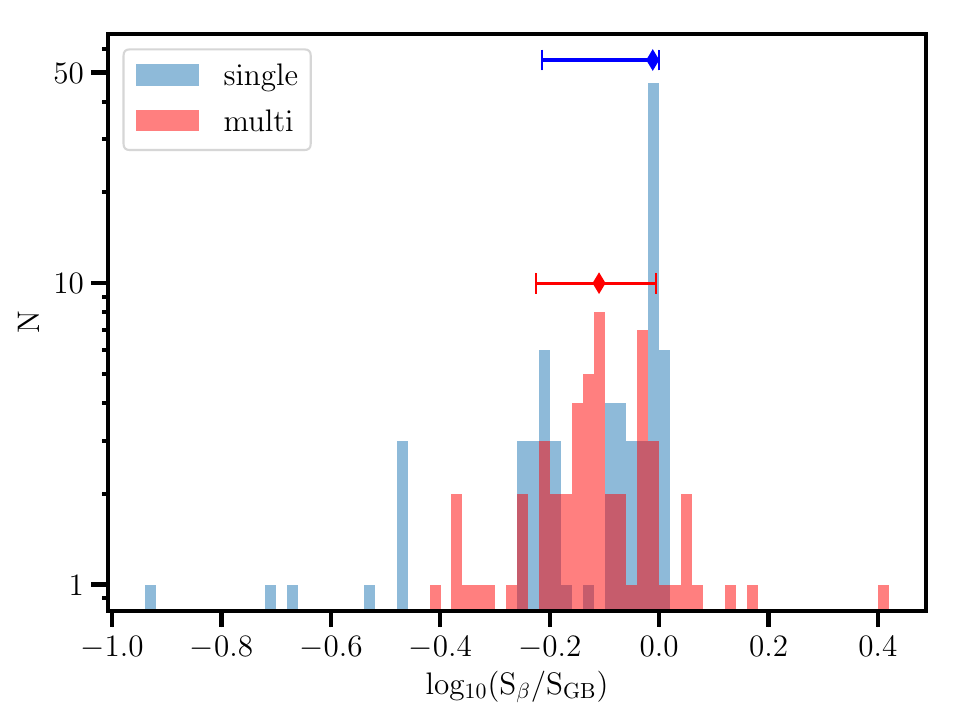}
    \caption{The distribution of the ratio between the power-law and gray-body (GB) predicted dust emission at rest-frame 850\,$\mu$m. The blue/red histogram considers the cases when one/two photometry data are available. Note that axis are in logarithmic scale. The errorbars show the 16$^{\rm th}$, 50$^{\rm th}$ (median), and 84$^{\rm th}$ percentiles.}
    \label{fig:gbvsspid}
\end{figure}

Figure~\ref{fig:gbvsspid} shows the distribution of the ratio between the power-law and GB predicted dust emission at rest-frame 850\,$\mu$m. Two histograms are displayed for cases when only one or two photometry data are available (blue and red, respectively). The blue histogram (representing most of our cases) very much peaks at a ratio of $\sim$1 with a long tail extending to low values (meaning a larger value predicted by the GB assumption), with a median (16$^{\rm th}$ and 84$^{\rm th}$ percentiles) of $\log_{10}(S_\beta/S_{\rm GB})=-0.012$ (-0.21 and -0.00040). The same statistics are found to be -0.11 (-0.23 and -0.0058) for the red distribution. The latter is mostly related to the more curved nature of the GB model with respect to the power-law model, and the two-data-point fitting being worst for the GB model (i.e., it always passes between the two points). Having these results, and the fact that a larger dust flux estimate leads to a larger predicted HI content (for a fixed molecular content), we choose to be conservative and adopt the power-law estimate.

\subsection{Converting high-J CO transitions to J:1-0}

Although CO~J:1-0 is the reference transition with which to estimate total M$_{\rm H_2}$ even up to high redshifts \citep[e.g.,][]{Hainline06,Carilli10,Harris10,Ivison10,Ivison11,Riechers11}, it is a common practice to target a higher-J transition (e.g., J$_{\rm up}=2,3,4$). This owes to the fact that they are brighter, less optically thick, but they are also found at frequencies where the dust continuum emission is also brighter. For that reason, one needs to consider the CO SLEDs to convert from the higher-J transition to the ground one. In this work, we adopt the line luminosity ratios reported for sub-millimeter galaxies in Table~2 in \citet{CarilliWalter13} for the samples assembled from \citet{Walter2011}, \citet{Saintonge2013}, PHIBSS, \citet{Birkin2021}, and any other galaxy in other samples at $z>2$ (see Section~\ref{sec:sample} for sample details). Otherwise, we adopt the line luminosity ratios reported for galaxies at $1.0<z<1.6$ in \citet{Boogaard2020}. We do this separation following the findings in \citet{Boogaard2020}, acknowledging that selection effects tend to provide samples with more excited CO SLEDs at higher redshifts. We do note that the CO SLEDS reported by \citet{Boogaard2020} at $1.0<z<1.6$ are in line with those by \citet{Daddi2015}, while the SMG CO SLEDs reported by \citet{CarilliWalter13} are in line with those by 
\citet{Bothwell13} and \citet{Birkin2021}, and slightly lower than those by \citet{Boogaard2020} at $2.0<z<2.7$.

\subsection{Corrections to a hotter CMB} \label{sec:cmb}

Due to the adiabatic expansion of the Universe, the Cosmic Microwave Background (CMB) shows today a black-body spectral energy distribution characteristic of temperature of 2.73\,K \citep{Fixsen09}, while, as we look back in time, the CMB shows a hotter temperature. This results in the molecular gas and dust becoming progressively in thermal equilibrium with the CMB, thus making it harder to detect emission against a brighter/hotter background emission \citep{daCunha13,Zhang16}. Here, we adopt the correction recipes proposed by \citet{daCunha13}, namely, we make use of Equations~12 and 18 therein.

\subsection{Estimating M$_{\rm HI}$}

In Equations~\ref{eq:mism} through ~\ref{eq:mismLR}, there are three main observables: $S_{850}$, $S_{HI}\Delta_v$, and $S_{CO}\Delta_v$. Thus, if one detects two of them, the third one can be inferred within the errors associated with both the photometric errors and the intrinsic population scatter shown in Figure~\ref{fig:dust2ism}. Nowadays, with the growing legacy of ALMA and NOEMA surveys, the most common scenario is when a (sub-)mm facility observes both $S_{850}$ and $S_{CO}$, sometimes in a single observation. As a result, in a statistical sense, one can attempt to derive the HI content in a galaxy population.

We thus combine Equation~\ref{eq:mism} with Equations~\ref{eq:mismRT} and \ref{eq:mismLR} to determine M$_{\rm HI}$:

\begin{equation}
    {\rm M_{HI}^{RT} = \frac{10^{\log(L_{850}[W/Hz]) - 12.54}}{\alpha_{heavy}} - 1.21~L'_{CO10} }
    \label{eq:mhiRT}
\end{equation}

\begin{equation}
    {\rm M_{HI}^{LR} = \frac{10^{(\log(L_{850}[W/Hz]) - 14.14)/0.823}}{\alpha_{heavy}} - 1.21~L'_{CO10} }
    \label{eq:mhiLR}
\end{equation}

We note that in cases where the derived M$_{\rm HI}$ is negative, we interpret such a result as a galaxy whose ISM gas component is mostly in its molecular phase.

\subsection{Error budget} \label{sec:error}

As one can see from the previous sub-sections, there are different steps involved up until when one retrieves the final HI mass estimate. In this section, we summarize the different uncertainties that are being considered in quadrature to retrieve the final error budget associated to the reported HI mass value:
\begin{itemize}
    \item we have adopted the {\sc mad} values associated with the adopted relations (Section~\ref{sec:dust2ism})
    \item we have considered population {\sc std} associated to the reported luminosity ratios between higher-J and J:1-0 transitions from \citet{Boogaard2020}, while we have adopted a general error of 0.15 associated to the values reported by \citet{CarilliWalter13}, since no error is reported therein, but it is nevertheless in line with other works \citep[e.g.,][]{Boogaard2020,Birkin2021}
    \item to the reported CO and continuum errors, we have added 10\% of the flux in quadrature to account for absolute flux scaling systematics
    \item for those galaxies with only one photometry data point in the mm spectral range, we have considered the {\sc mad} associated to the adopted spectral index.
    \item in the luminosity functions and cosmological mass density estimates, we account for cosmic variance following \citet{TrentiStiavelli08}. We use their calculator\footnote{Version v1.03: https://www.ph.unimelb.edu.au/$\sim$mtrenti/cvc/CosmicVariance.html} to compute the error fraction associated to cosmic variance based on redshift range, field size, and number of sources. Based on this exercise, we adopt an error fraction of 0.25 and 0.30 for B21 and ASPECS samples, respectively, to be added in quadrature.
\end{itemize}

\iffalse
% cosmic variance
% https://www.ph.unimelb.edu.au/~mtrenti/cvc/CosmicVariance.html
- intrinsic number of objects: 1 / 6 / 16 / [20 ASPECS / 30 B21]
- completeness: 0.8
- halo filling factor = 1
- redshit: 1.4±0.4; 2.6±0.6
- area [arcmin2]: 76x76 (cosmos); 59x59 (uds); 2.1x2.1 (ASPECS)
- OmLamb=0.74, OmMat=0.26, H=70, specIdx=1, sig8=0.9, bias=Press-Schechter

cosmos: lowz 0.35/0.26/0.23; highz 0.25/0.2/0.18
uds: lz 0.39/0.29/0.25; hz 0.29/0.23/0.2
aspecs: lz 0.35/0.24/0.20; hz 0.43/0.31/0.25
\fi

%% ==== SAMPLE SELECTION
\section{Sample selection} \label{sec:sample}

As stated in the previous section, in this manuscript, we make use of direct observations of mm-continuum and CO to infer the content of HI gas. As a result, we have assembled from the literature a sample of galaxies that have both a CO emission line detection and (sub-)mm continuum coverage. We mainly choose galaxies with a low-J$_{\rm up}$ CO transition (typically J$_{\rm up}\leq3$) in order to have a more reliable conversion to CO J:1-0.

As you will see ahead, specifically for the analysis of the cosmic HI mass content (Section~\ref{sec:mfdens}), we focus only on two samples (Sections~\ref{sec:aspecs} and ~\ref{sec:b21flux}) which provide the simplest selection function.

\subsection{ASPECS} \label{sec:aspecs}
We consider galaxies from the ALMA SPECtroscopic Survey in the Hubble Ultra-Deep Field (ASPECS) LP survey \citep{GonzalezLopez2020,Walter2016,Aravena2016} later followed up at lower frequencies by its VLA equivalent VLASPECS \citep{Riechers20}. Both programs comprise blind surveys of CO covering an area of 4.6 arcmin$^2$. We focus on the CO~J$_{\rm up}=1,2,3$ line measurements from \cite{Boogaard2020} for 18 galaxies. For those galaxies, we use the 1.2\,mm continuum flux densities from \cite{Aravena2020} and \cite{GonzalezLopez2020}, and the 3\,mm flux densities from \cite{GonzalezLopez2019}. We also looked for other sub-mm observations and found a 870\,$\mu$m flux density measurement for the galaxy ``1mm6'' in the \cite{Chapin2011} catalog.
 
\subsection{Birkin et al. (2021)} \label{sec:b21flux}
\cite{Birkin2021} presented ALMA and NOEMA observations of SMGs selected from the ALMA-SCUBA-2 Cosmic Evolution Survey (AS2COSMOS, total area of 1.6 deg$^2$; \cite{Simpson2020}), the ALMA-SCUBA-2 Ultra Deep Survey (AS2UDS, total area of 0.96 deg$^2$; \cite{Stach2019,Dudzeviciute2020}) and the ALMA-LABOCA ECDFS Submillimetre Survey (ALESS, area of 30’ × 30’; \cite{Hodge2013,Danielson2017}). The original catalog provides the 870$\mu$m and 3mm flux densities for 61 galaxies, along with CO emission lines, J$_{\mathrm{up}}$ = 2--5 for 50 of those galaxies.

For the HI mass density analysis in Section~\ref{sec:mfdens}, we focus on the flux-selected galaxies within the ``scan sample'' in \citet{Birkin2021}: 5 source with $S_{870}=15-20\,$mJy in AS2COSMOS and 13 sources with $S_{870}=8-14\,$mJy in AS2UDS. Note the complementary flux selection. In additon, we also considered the 13 optical/near-infrared faint galaxies within the ``scan sample'' and the 30 galaxies within the ``$Spec-z$ sample''.

%depth: typical RMS depth of 0.3 mJy for ALMA, typical RMS depth of 0.7 mJy for NOEMA

\subsection{COLDz}
The CO Luminosity Density at High-z (COLDz; \cite{Pavesi2018}) survey is a blind survey of CO that covered $\sim$ 9 arcmin$^2$ of the COSMOS deep field and $\sim$ 51 arcmin$^2$ of the GOODS-North wide field.
We found the sub-mm photometry in different surveys. For the galaxies COLDz.GN.31 and COLDz.COS.0 we found the 850$\mu$m flux density from the SCUBA-2 Cosmology Legacy Survey \cite{Geach2017}; for the galaxies COLDz.GN.14 and COLDz.GN.16 we use the 850$\mu$m flux densities from the SUPER GOODS survey \cite{Cowie2017}; for the galaxy COLDz.COS.11 we use the 1100$\mu$m flux density from the AzTEC millimetre survey of the COSMOS field \cite{Aretxaga2011}; and finally, using both the \cite{Liu2018} and \cite{Jin2018} catalogs we adopted the 850$\mu$m photometry for the galaxies COLDz.GN.15, COLDz.GN.28, COLDz.COS.6, and both the 850$\mu$m and 1100$\mu$m photometry for the galaxies COLDz.GN.0 COLDz.GN.3,COLDz.COS.1, COLDz.COS.2, COLDz.COS.3. 
This resulted in 13 selected galaxies. 

%\subsection{Representative sample}

\subsection{PHIBSS}
The IRAM Plateau de Bure HIgh-z Blue Sequence Survey (PHIBSS; \citealt{Tacconi13}) observed the CO\,(3-2) line emission for 52 massive, main-sequence star-forming galaxies. These galaxies were chosen from UV/optical/IR surveys to study the molecular gas in galaxies near the cosmic star formation peak of normal galaxies, and were carefully selected to  cover a complete M$_*$-SFR plane.
They included two redshift bins, at $z\sim1.1$ and $\sim$2.2. The first bin, z = 1-1.5 includes galaxies from the All-Wavelength Extended Groth Strip
International Survey (AEGIS; \citealt{Davis2007}), which includes imaging from X-ray to radio and optical spectroscopy.
The higher redshift bin, z = 2-2.5, includes galaxies from \cite{Erb2006}, \cite{Mancini2011}, the \emph{BzK} sample of \cite{Daddi2010}, \cite{Magnelli2012} and the three lensed galaxies cB58 \citep{Baker2004}, ``cosmic eye'' \citep{Coppin2007} and ``eyelash'' \citep[][also known as J2135-0102]{Swinbank2010}. 
We found sub-mm observations for some of these galaxies in the following catalogs/surveys. We include 850$\mu$m fluxes from the main catalog of the SCUBA-2 Cosmology Legacy Survey \citep{Geach2017} for the galaxy EGS13004291 and the fluxes for the catalog of the EGS deep field \citep{Zavala2017} for the galaxies EGS13011155, EGS13011166, EGS13017707, EGS13018076. For the $BzK$ galaxies we include the 1300$\mu$m fluxes from \cite{Magdis2012}. From the \cite{Liu2018} catalog we include the 850$\mu$m flux for the BzK17999 galaxy and both the 850$\mu$m and 1100$\mu$m fluxes for the galaxies BzK12591 and PEPJ123633. For the lensed galaxy ``eyelash'', we include the 850$\mu$m and 1200$\mu$m fluxes from \cite{Ivison2010}; for the galaxy cB58 we include the 850$\mu$m flux from \cite{vanderWerf2001}; for ``cosmic eye'' we include the 1200$\mu$m flux from \cite{Saintonge2013}; and for the galaxy Q1700-MD94 we include the 1200$\mu$m flux from \cite{Henriquez-Brocal2022}.
This results in 15 galaxies with sub-mm measurements %and 8 that were included using an SED fit with Herschel information, with a total of 23 galaxies 
from the PHIBSS sample.

\subsection{PHIBSS2}

The PHIBSS2 survey \citep{Freundlich2019} is an extension of the PHIBSS sample, previously described.
In this survey, the CO(2-1) line emission was detected for 60 normal star-forming galaxies at redshifts $z = 0.5-0.8$. These galaxies were drawn from the North field of the Great Observatories Origins Deep Survey \citep[GOODS-N; ][]{Giavalisco04}, the Cosmic Evolution Survey \citep[COSMOS; ][]{Scoville07}, and the AEGIS survey \citep{Davis07}. They were chosen because deep $HST$, good quality spectroscopy and UV/IR observations are available, while following similar selection criteria as in PHIBSS.

The sub-mm information was gathered from the following surveys/catalogs. 
For the galaxies XG55 and L14GN022 we used the 850$\mu$m fluxes from the SCUBA-2 Cosmology Legacy Survey \citep{Geach2017}; for the galaxies XC54, XF54 and L14EG008 we used the 850$\mu$m fluxes from the EGS deep field sample of the SCUBA-2 Cosmology Legacy Survey \citep{Zavala2017}; for the galaxy L14GN034 we used the 850$\mu$m flux from the SUPER GOODS survey \cite{Cowie2017}. From the \cite{Liu2018} catalog we include the 850$\mu$m and 1100$\mu$m fluxes for the galaxy XA55.
This results in 7 galaxies with sub-mm measurements.%, plus 5 with the estimated submm through a fit. 

\subsection{Saintonge et al. (2013)}
\cite{Saintonge2013} compiled a sample of 17 lensed galaxies to study dust and gas at the high redshifts of $z = 1.4-3.1$. This goal was achieved combining observations in the FIR with $Herschel$ PACS/SPIRE, and of the CO\,(3-2) line emission, using the IRAM Plateau de Bure Interferometer (PdBI). Also, they include the 1.2mm continuum photometry from IRAM 30m, Max Planck Millimeter Bolometer array (MAMBO; \citealt{Kreysa1998}) and Submillimeter Array (SMA) observations.
These galaxies are considered to be UV-bright lenses, that show similar SFRs and stellar masses to main-sequence galaxies at high redshift, thus, they are normal star-forming galaxies.

We selected 5 lensed galaxies, that are the detected CO\,(3-2) line. %Some of the galaxies from the original catalog were discarded due to large errors in the FIR photometry. Plus, 
We discarded the sources cB58, Eye and Eyelash since they are part of the PHIBSS sample. 

\subsection{Seko et al. (2016)}

The sample from \cite{Seko2016} is extracted from a larger one reported in \citet{Yabe12} where 317 sources selected from optical and NIR surveys (SXDS/UDS; 0.67~deg$^2$) were followed-up spectroscopically, with 71 being assigned a reliable redshift estimate. Of those, 20 were targeted by \cite{Seko2016}, with 11 being detected in CO~(5-4), of which 5 were detected in mm continuum.

\subsection{VALES}
The Valpara\'iso ALMA Line Emission Survey (VALES; \citealt{Villanueva17}), used ALMA Band 3 to observe the CO(1-0) line emission in 67 galaxies in the redshift range 0.02 < z < 0.35. This survey targeted galaxies from the Herschel Astrophysical Terahertz Large Area Survey (H-ATLAS; \citealt{Eales2010}), that covered a total area of $\sim$160 deg$^2$ of the sky \citep{Valiante2016}. These galaxies were selected because they are detected near the peak of the SED for a normal, local and dusty star-forming galaxy \citep{Villanueva17}. The sample have reliable matches with the 6$^{\rm th}$ Sloan Digital Sky Survey data release \citep[SDSS;][]{Adelman-McCarthy08}. The sample originates from two different selections involving criteria such as SDSS sizes or \textit{Herschel} PACS spectroscopy.

Out of the 67 sources, 49 show a CO\,(1-0) line detection (>$5\sigma$). For 25 of these galaxies we found the 1\,mm continuum using the ALMA archive, from the projects 2016.1.00994.S, 2017.1.01647.S, and 2017.1.00287.S. These observations are down to a depth of $\sim$0.02 mJy/beam, $\sim$0.003 mJy/beam and $\sim$0.03 mJy/beam, respectively. %For the other 24 galaxies, we use the \emph{Herschel} FIR photometry and derive the 850$\mu$m with a fit. This results in a total of 53 sources.

\subsection{Walter et al. (2011)}
\cite{Walter2011} compiled 850$\mu$m, CO(3-2) and CI observations for different SMG galaxies. We selected 15 of their catalog galaxies for our sample, while the remainder were discarded due to the uncertainty on the magnification factor. 

The sub-mm continuum flux information comes from:
\citet[][SMM J02399-0136]{Ivison1998}, 
\citet[][BRI 1335-0417]{Benford1999},
\citet[][SMM J14011+0252]{Ivison2000}, 
\citet[][RX J0911+0551]{Barvainis2002}, 
\citet[][F10214, Cloverleaf]{Barvainis2002}, 
\citet[][PSS J2322+1944]{Isaak2002}, 
\citet[][SMM J163650+4057, SMM J163658+4105]{Ivison2002}, 
\citet[][SMM J123549+6215]{Chapman2003}, 
\citet[][SMM J16359+6612]{Kneib2004}, 
\citet[][SDSS J1148+5251]{Robson2004}, 
\citet[][GN 20, GN 20.2]{Pope2006}

\subsection{Summary}

Table \ref{tab:sample-summary} provides an overall perspective of each survey or sample considered here. It includes the number of sources, the range in redshift and, when appropriate, the survey areal size. The names in the first column are the aliases by which we will refer to the samples henceforth. Overall, we have assembled a total of 335 galaxies, selected in different wavelength regimes, from optical continuum or spectroscopy, to near-infrared and (sub-)millimeter. As mentioned already, the samples ASPECS and B21 are the samples providing the simplest selection function hence, these are the two samples used to compute the cosmic HI mass density content in Section~\ref{sec:mfdens}.

The full sample is provided in a master table as supplementary material.

\begin{table}
\caption{Analysis Sample Summary --- this table reports the different samples we have considered to retrieve HI mass content in this work. The first column reports the aliases used throughout the manuscript. The third to fifth columns report the redshift ranges each sample covers, their surveyed areas, and the number of galaxies in each one.} 
\centering
\footnotesize
\begin{tabular}{ccccc}  \hline
Sample & Reference & Redshift & Area & N   \\ \hline
VALES & \cite{Villanueva17}& 0.01--0.35 & 160\,deg$^2$ & 49 \\
PHIBSS & \cite{Tacconi13} & 1.0--2.3 & 0.41\,deg$^2$ & 23 \\
PHIBSS2 & \cite{Freundlich2019} & 0.50--0.78 & --- & 60 \\
S13 & \cite{Saintonge2013} & 1.4--2.7 & --- & 5 \\
W11 & \cite{Walter2011} & 2.2--6.4 & --- & 14 \\
ASPECS & \cite{Walter2016} & 0.5--2.7 & 4.6\,arcmin$^2$ & 20 \\
COLDz & \cite{Pavesi2018} & 2.0--5.3 & 60\,arcmin$^2$ & 58 \\
S16 & \cite{Seko2016} & 1.3--1.6 & 0.77\,deg$^2$ & 11 \\
B21 & \cite{Birkin2021} & 1.2--4.8 & 2.81\,deg$^2$ & 50 \\  \hline
&  & & \textbf{Total} & 335 \\ 
\hline  
\end{tabular}
\label{tab:sample-summary}
\end{table}

%% ==== RESULTS
\section{Results} \label{sec:results}

\subsection{HI mass estimates}

\begin{figure}
    \centering
    \includegraphics[width=0.48\textwidth]{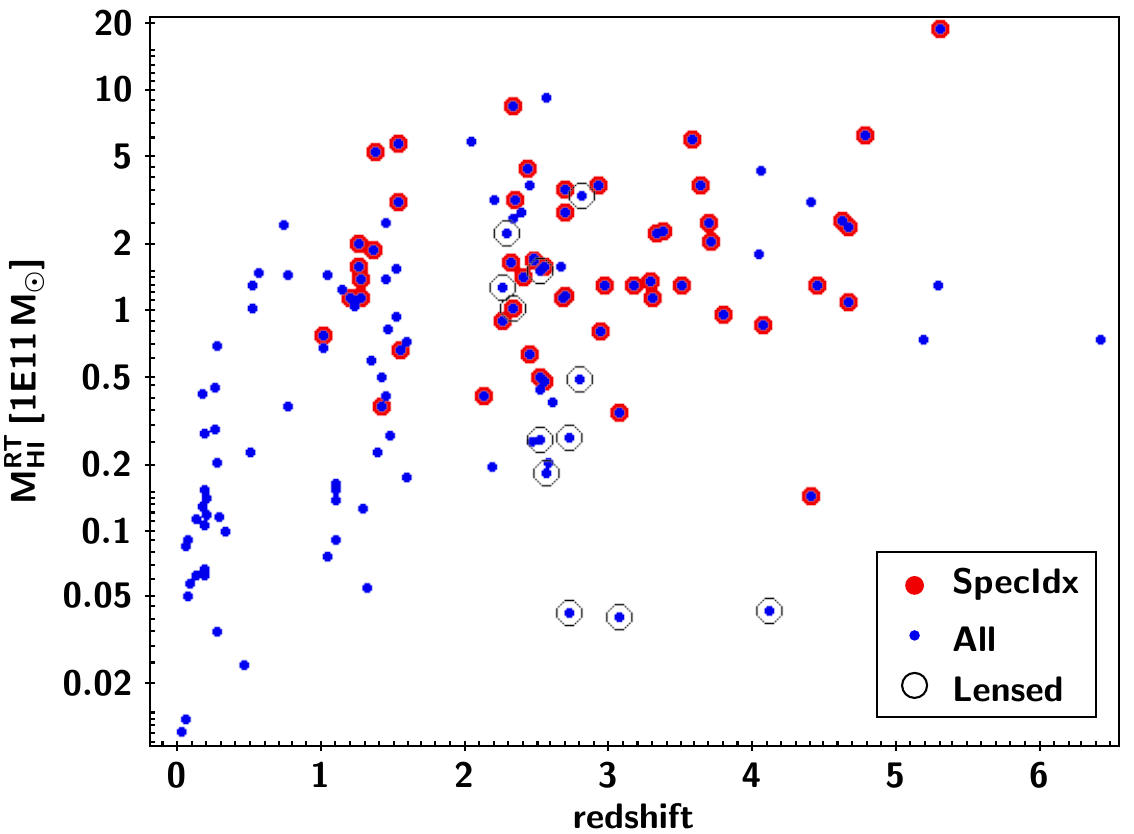}
    \caption{The M$^{\rm RT}_{\rm HI}$ distribution with redshift for 149 with non-zero and non-negative M$_{\rm HI}$ estimates adopting the RT method. The values do not consider the contribution of heavier elements. Galaxies that have more than one photometry data points in the mm spectral range (labelled SpecIdx) are highlighted by red dots. Galaxies that are known lensed galaxies are highlighted by empty black circles. For the latter, the plotted values are corrected for gravitational magnification.}
    \label{fig:mhiz}
\end{figure}

\begin{figure}
    \centering
    \includegraphics[width=0.48\textwidth]{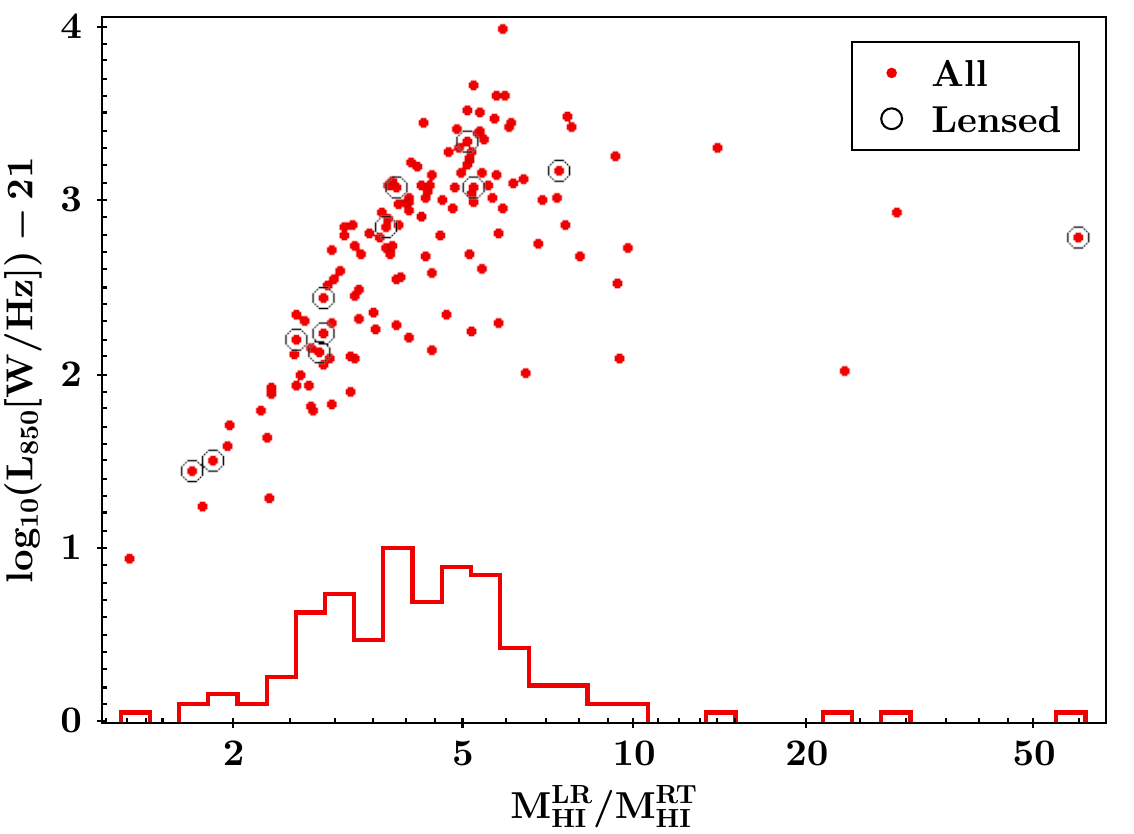}
    \caption{The ratio between the M$_{\rm HI}$ estimates derived with LR and RT methods versus $L_{850}$. A total of 137 galaxies, for which both methods predict non-zero/negative values, are used in this plot. For reference, the lensed galaxies are highlighted with black circles. The inset histogram is a normalized one to show the distribution of the data set. The distribution shows that LR predicts a median factor of $\sim$4 higher M$_{\rm HI}$ content than the RT method. The \textit{y}-axis was normalized to $10^{21}$ for display better visual comparison between the scatter plot and the histogram.}
    \label{fig:methodcomp}
\end{figure}

We derive non-zero and non-negative M$_{\rm HI}$ estimates for 149 and 145 galaxies while using the RT or LR methods (Section~\ref{sec:method}), respectively. In Figure~\ref{fig:mhiz}, we show the M$^{\rm RT}_{\rm HI}$ distribution with redshift. We highlight those galaxies that have more than one photometry data points in the mm spectral range (red dots), and those which are known lensed galaxies (empty black circles; magnification-corrected values are displayed). There are no clear deviations between these groups, except for a few lensed galaxies understandably showing lower masses. If one instead plots M$^{\rm LR}_{\rm HI}$ the data points will shift to higher M$_{\rm HI}$ values by a median factor of 4.3 (with a spread of 0.18\,dex or a factor of 1.5) as shown in Figure~\ref{fig:methodcomp} (see also Section~\ref{sec:gasfrac}). This is expected given the flatter slope of the LR estimate, that implies a higher M$_{\rm HI}$ component for the same $L_{850}$ and $L'_{\rm CO}$ values with respect to RT.

\subsection{ISM gas mass fractions} \label{sec:gasfrac}

\begin{figure}
    \centering
    \includegraphics[width=0.49\textwidth]{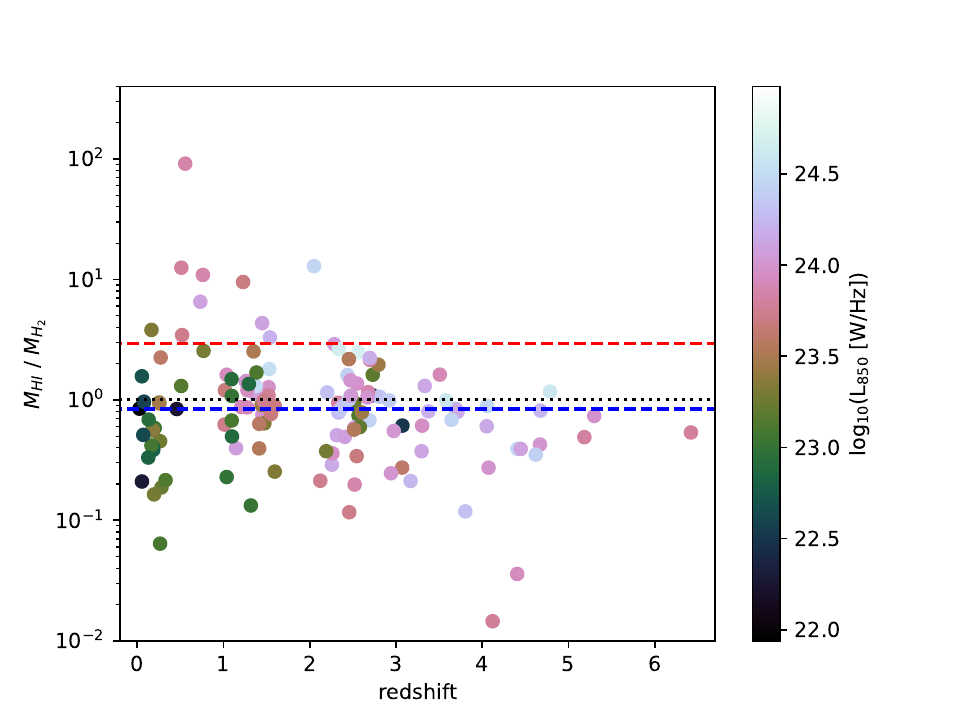} \\
    \includegraphics[width=0.49\textwidth]{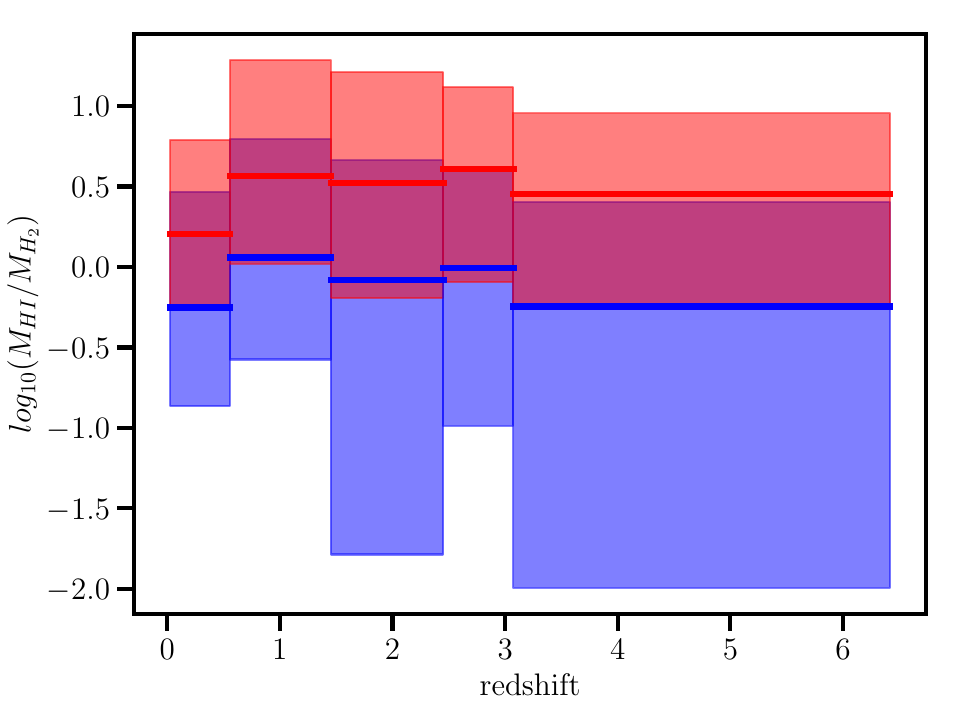} \\
    \includegraphics[width=0.49\textwidth]{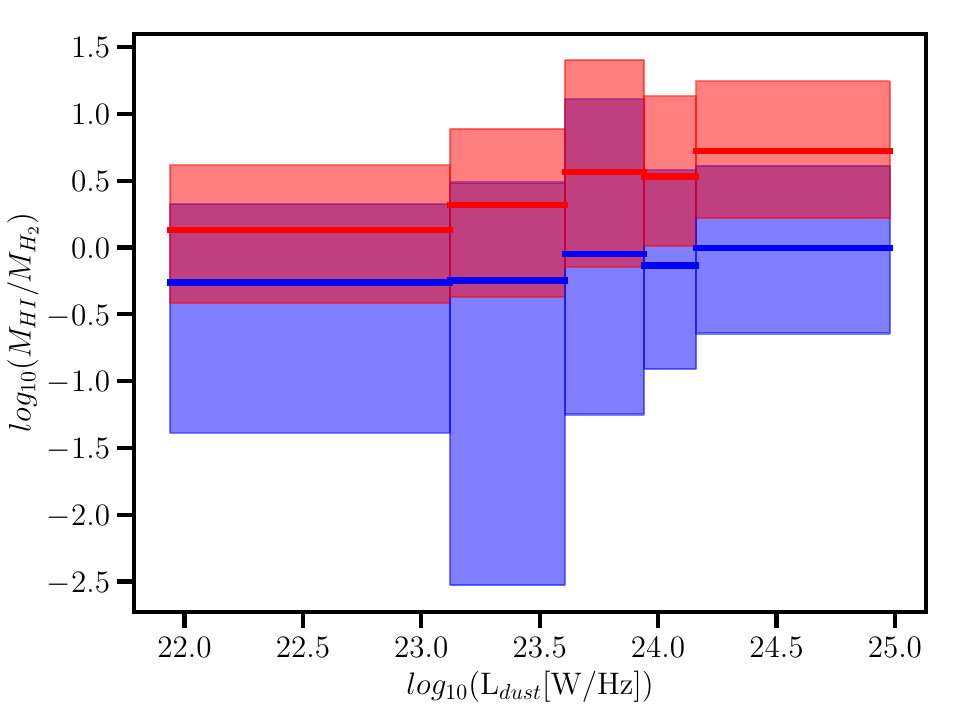}
    \caption{Distribution of M$_{\rm HI}$/M$_{\rm H_2}$ ratio with redshift and $L_{850}$. \textit{Top panel}: The data points show the results adopting RT. The color coding reflects $L_{850}$. For reference, we mark the equal-content level as a dotted-line, and the median values for RT (dashed blue) and LR (dashed red) methods. \textit{Middle panel}: The dependency of M$_{\rm HI}$/M$_{\rm H_2}$ ratio on redshift separating the sample into quantiles. The thick lines show the median value in each quantile (red and blue lines for LR and RT methods, respectively). Each box shows where 68\% of the population falls (16$^{\rm th}$ and 84$^{\rm th}$ percentiles) based on a Bootstrapping analysis (see details in the text). \textit{Bottom panel}: Same as in the middle panel, but with respect to $L_{850}$.}
    \label{fig:gasratios}
\end{figure}

We now compare the M$_{\rm HI}$ and the M$_{\rm H_2}$ content in our sample. This comparison is shown in Figure~\ref{fig:gasratios} for both RT and LR. For reference in the top panel, we mark the equal-content level as a dotted-line, and the median values for RT (dashed blue) and LR (dashed red) methods. The color coding reflects $L_{850}$. For reference, $L_{850}=10^{23}$ and $10^{24}$\,W/Hz are approximately the LIRG and ULIRG thresholds \citep[based on Equation E.5 in ][]{Orellana17}. The middle and bottom panels in Figure~\ref{fig:gasratios} show the dependency of the M$_{\rm HI}$/M$_{\rm H_2}$ ratio with redshift or $L_{850}$, respectively, for the RT and LR methods (black and red color, respectively). The sample was divided into quantiles, within which the median value is estimated (thick lines). The boxes show where 68\% of the population within each quantile falls. These ranges were estimated based on a Bootstrapping analysis, where, for each gas ratio estimate and associated error, we randomly draw 100 new values assuming a log-normal distribution. From these we then retrieve the 16$^{\rm th}$ and 84$^{\rm th}$ percentiles which delimit the boxes. Overall, there is no significant evidence for evolution of the M$_{\rm HI}$/M$_{\rm H_2}$ ratio with redshift or $L_{850}$, except if one adopts the LR method which may imply higher ratios with increasing $L_{850}$ (red trend in bottom panel). This is not unexpected, since it is at the highest luminosities that the two relations deviate more from each other (Figure~\ref{fig:dust2ism}).

We do note that \citet{Chowdhury22} reports M$_{\rm HI}$/M$_{\rm H_2}$ ratios of 2--5\footnote{One detail worth noting is that \citet{Chowdhury22} estimate the molecular gas content based on a relation dependent on stellar mass and specific star formation rate.} for a sample of galaxies with ${\rm M^* =1.03\pm0.24\times10^{10}~M_\odot}$ at $0.74<z<1.45$. Although this is in line with the overall value of 2.9 ($-0.42,+0.36\,$dex) using LR method, versus 0.8 ($-0.56,+0.33\,$dex) using RT, the galaxy population is different. For those galaxies in our analysis sample at $0.74<z<1.45$ that have reported stellar mass estimates in the literature, 95\% have $\sim$M$^*>10^{10.2}\,$M$_\odot$. Also, the gas mass fractions are sample-selection dependent%, and the sample we use here does not have a homogeneous method which was used to estimate the stellar mass of the sample used here
. Despite the ratio values just mentioned for the analysis sample assembled in this work, the ASPECS and B21 samples used in Section~\ref{sec:mfdens} show lower values with different deviations in each redshift bin. Namely, the M$_{\rm HI}$/M$_{\rm H_2}$ ratio is 0.9 at $z\sim1.5$ and 0.5 at $z\sim2.5$ when using the RT method, and 2.4 and 2.8 when using LR. This goes in line with the fact that these samples are truly dust- and CO-selected (as opposed to other samples that are, e.g., selected in the optical-NIR), hence more likely to yield a larger molecular gas fraction (see next Section~\ref{sec:mfdens}.

\subsection{HI mass functions and overall gas density} \label{sec:mfdens}

In this section, we use only the samples from ASPECS and B21 since these are the samples with the simplest selection function, because it only depends on the continuum and CO selections. In the B21 sample, we only consider the sub-sample referred to as ``scan sample'' (point i-a) therein. We decided not to use the COLDz sample, since the catalogue includes lower fidelity line detections, but the fidelity values are not reported. ASPECS and B21 spectral scans cover similar redshift ranges of interest to this work: $1.01<z<1.81$ and $2.01<z<3.20$. The samples are quite complementary, where ASPECS covers better the M$_{\rm H_2}<6.5\times10^{10}$M$_\odot$ regime, and B21 otherwise. We used this M$_{\rm H_2}$ value as threshold above/below which we considered the B21/ASPECS samples. We do this to prevent double counting sources with M$_{\rm H_2}$ values around that threshold while using the $V_{max}$ approach (see next paragraph), especially those above the adopted threshold present in ASPECS since cosmic variance may become critical. Nevertheless, Appendix~\ref{app:overlap} does show the implications of not adopting this mitigation approach.

In order to determine the volumetric representativeness of each galaxy we adopt the $V_{max}$ method. In order to determine the minimum and maximum redshifts ($z_{min}$ and $z_{max}$), we consider both the continuum and CO fluxes.
The minimum redshift is basically given by the limits of the spectral scans adopted in each survey.
To estimate $z_{max}$, we not only consider the significance of the flux in the detection band and the estimated or adopted spectral-index, but also the significance of the CO detection.

\begin{figure*}
    \centering
    \includegraphics[width=0.49\textwidth]{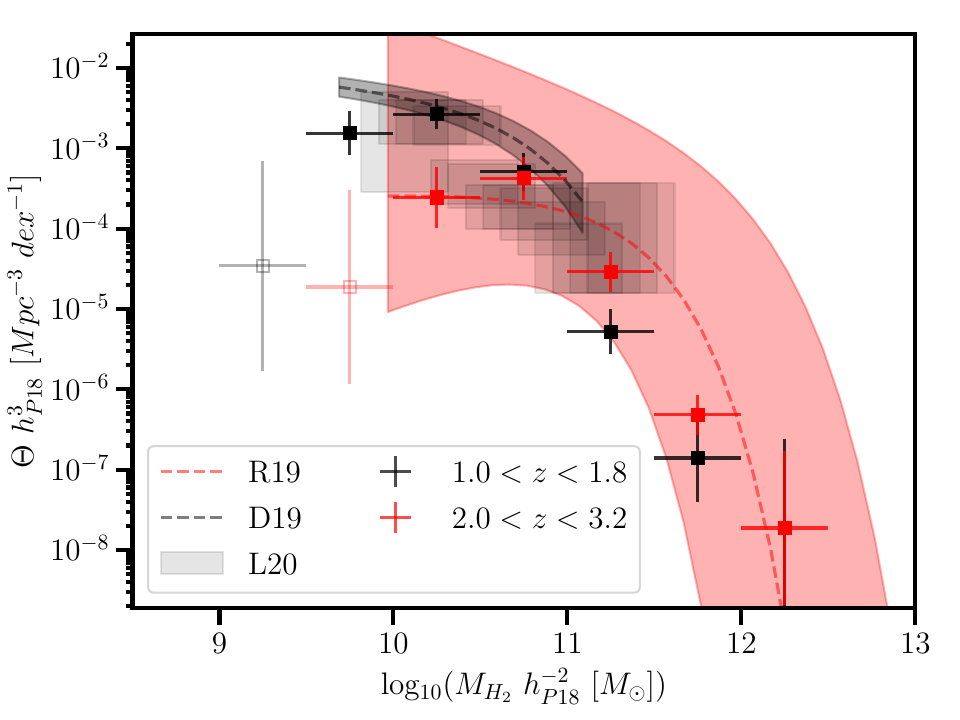}
    \includegraphics[width=0.49\textwidth]{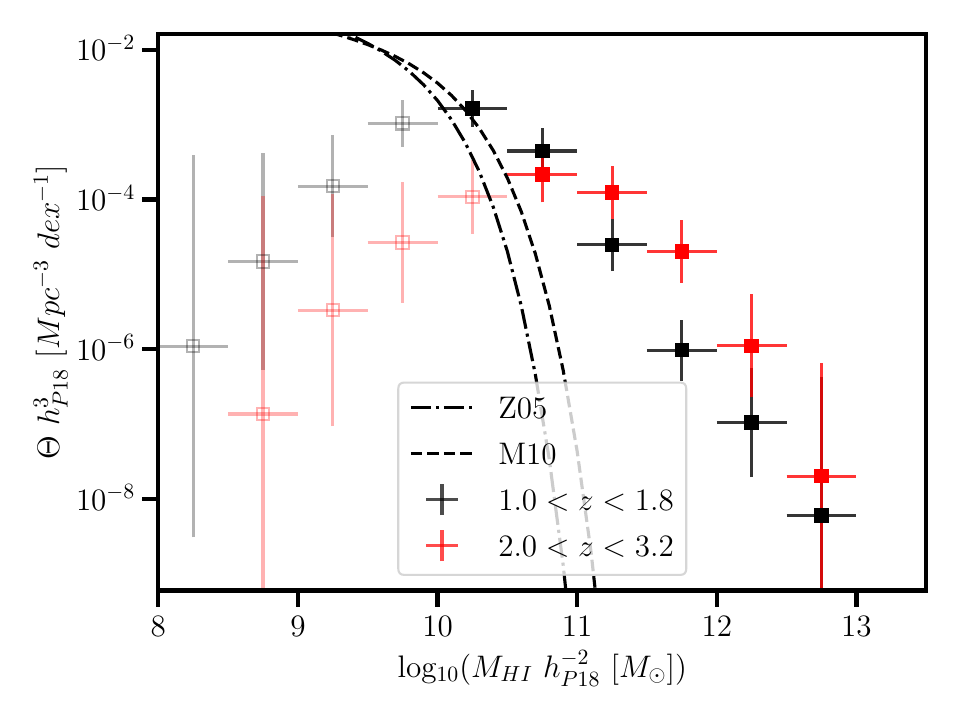}\\
    \includegraphics[width=0.48\textwidth]{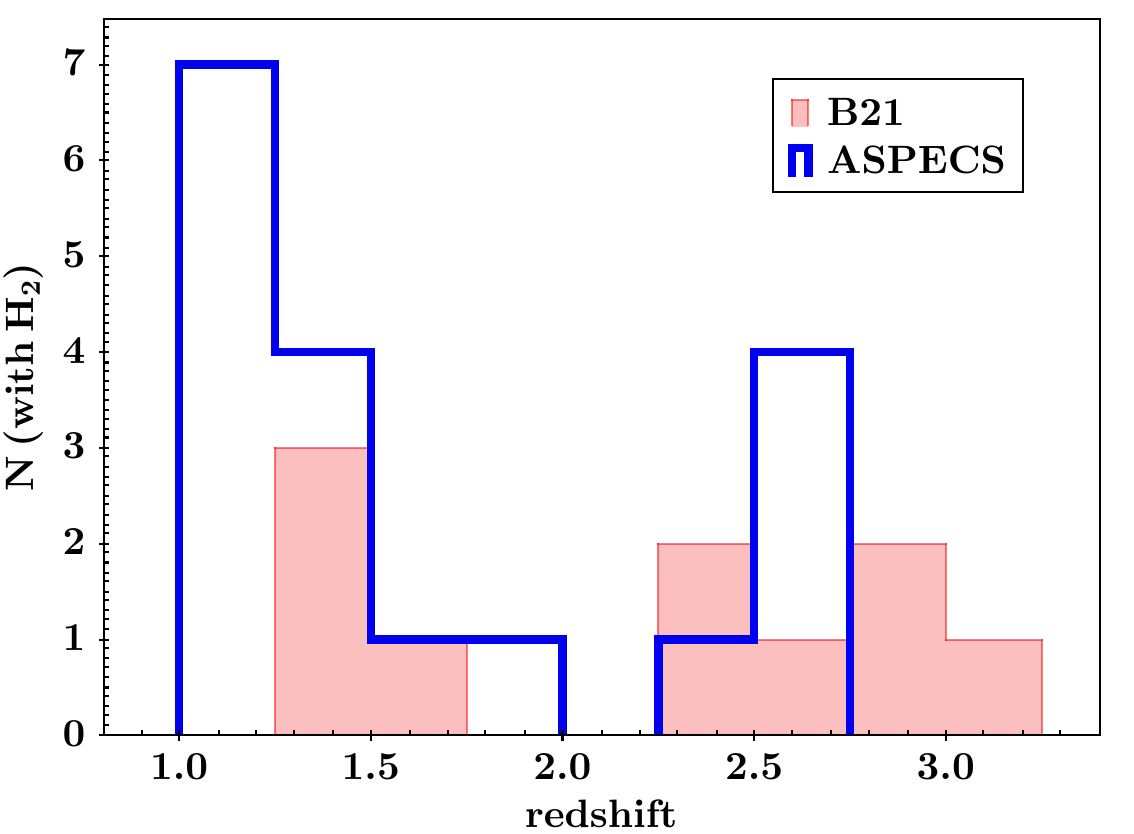}
    \includegraphics[width=0.48\textwidth]{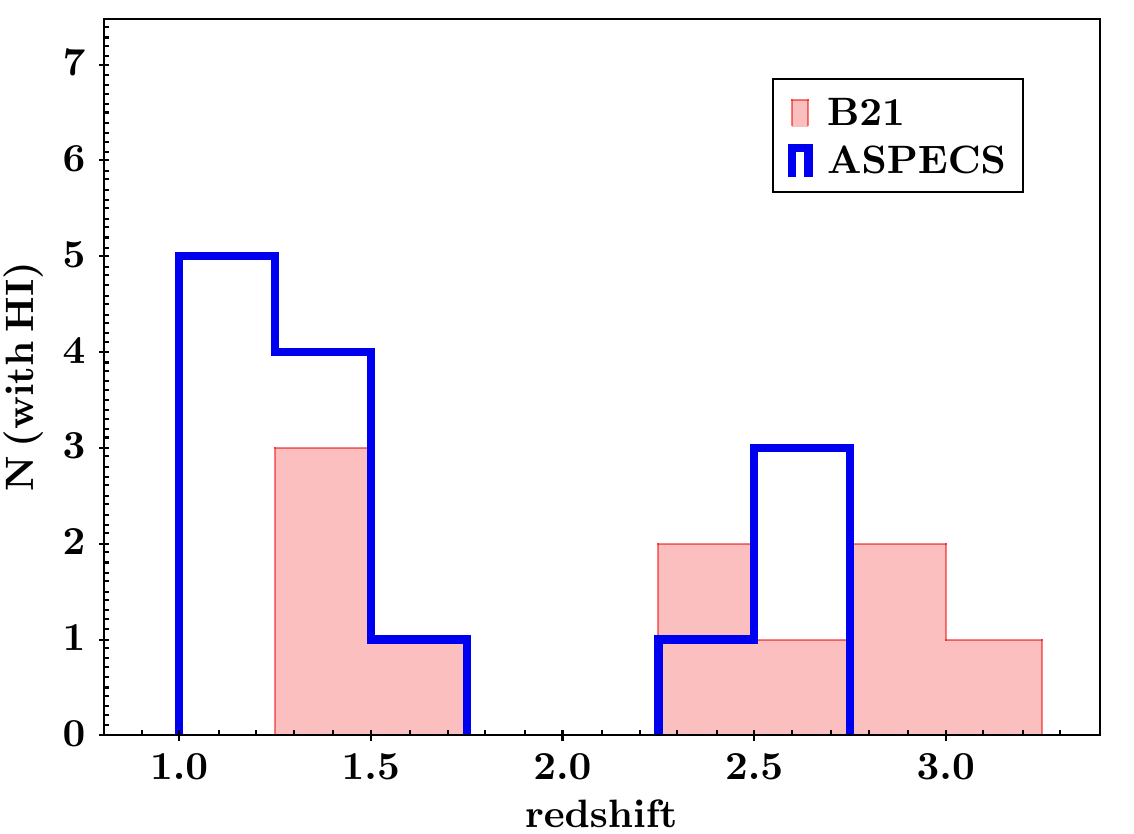}
    \caption{The mass functions (MFs; top row) for M$_{\rm H_2}$ (left-hand side panel) and M$_{\rm HI}$ (right panel; results for the RT method; see Appendix~\ref{app:lr} for results with the LR method) obtained with the ASPECS and B21 samples. Black and red data points refer, respectively, to the following redshift bins: $1.01<z<1.81$, $2.01<z<3.20$. The literature ${\rm H_2}$ MFs come from \citet[][dashed black line and gray shaded region; $<z>=1.43$]{Decarli19}, \citet[][gray boxes; $z\sim1-2$]{Lenkic20}, and \citet[][red dashed line and shaded region; $2<z<3$]{Riechers19}. The literature M$_{\rm HI}$ MFs are those reported in the local Universe by \citet[][dashed-dotted line]{Zwaan05}, \citet[][dashed line; \citet{Jones18} results match this one]{Martin10}. Translucent data points are indicative of the mass thresholds below which the MFs are believed to be incomplete. The bottom row reports the redshift distributions for ASPECS (blue histogram) and B21 (red shaded region) samples used in this analysis. The left-hand side panel shows the distribution of sources with M$_{\rm H_2}$ estimates, and the right panel shows those with positive M$_{\rm HI}$ estimates with the RT method.}
    \label{fig:massfuncs}
\end{figure*}

In Figure~\ref{fig:massfuncs}, we show the derived gas mass functions (MFs) for both M$_{\rm H_2}$ (top left-hand panel) and M$_{\rm HI}$ (top right; adopting the RT method, while Appendix~\ref{app:lr} shows the LR method) and compare them with estimates from the literature. All the M$_{\rm H_2}$ measurements from the literature have been converted to the $\alpha_{CO}$ value adopted in this work, and, once more, we do not correct for heavier elements. The agreement with the literature %at $2<z<3$ \citep{Riechers19}, while at $1<z<2$ our MFs, especially at the light-end (M$^*<10^{11}\,$M$_\odot$), are at the lower-limit estimates 
\citep{Decarli19,Riechers19,Decarli20,Lenkic20} is noticeable. %The latter is likely related to our approach of not applying any further completeness correction factor as done, for instance, in \citet{Decarli19,Decarli20}. 
%Nevertheless, the reader will see ahead that this does not affect the overall mass density estimates comparison with the literature.

The top right-hand panel shows the results for M$_{\rm HI}$ using the RT method. The local-Universe MFs from \citet{Zwaan05,Martin10,Jones18} are also displayed for reference (all limited to ${\rm M_{HI}<10^{11}\,M_\odot}$). Since the uncertainties in the derived M$_{\rm HI}$ estimate do spread over more than one bin (we adopted a standard bin width of 0.5\,dex), we actually distribute the measurement probability per bin assuming a log-normal error distribution since otherwise the massive-end would be underestimated \citep[see][we note that we have also adopted this approach while retrieving the ${\rm H_2}$ MFs]{Bera22}. Nevertheless, this approach also results in non-zero probability for bins that we believe refer to unrealistic gas masses (see Section~\ref{sec:rtvslr}). There is thus the concern that the high-mass end may be flatter than reality, while the normalization may be lower than reality. In Appendix~\ref{app:nopdflf}, we show that we do not find a significant trend supporting this expectation. Nevertheless, in line with this exercise, we discard mass bins at ${\rm M_{HI}>10^{12.5}\,M_\odot}$ for the RT method, and at ${\rm M_{HI}>10^{13}\,M_\odot}$ for the LR method. We do note that the peaks of distribution may not relate to the intrinsic shape of the functions, but may instead show the mass limit below which each sample is complete at each redshift range. Conservatively, we have thus made the data points more translucent below those thresholds. The thresholds adopted in the ${\rm H_2}$ MFs are those from the literature, while those in the ${\rm H_1}$ MFs are where one sees the trend inflex (+0.5 and +1\,dex higher for RT and LR methods, respectively, with respect to ${\rm H_2}$ MFs). Nevertheless, the result seems to point that the main difference in HI content with respect to the local Universe happens in galaxies with M$_{\rm HI}\gtrsim10^{10.5}\,$M$_\odot$.

\begin{figure*}
    \centering
    \includegraphics[width=0.49\textwidth]{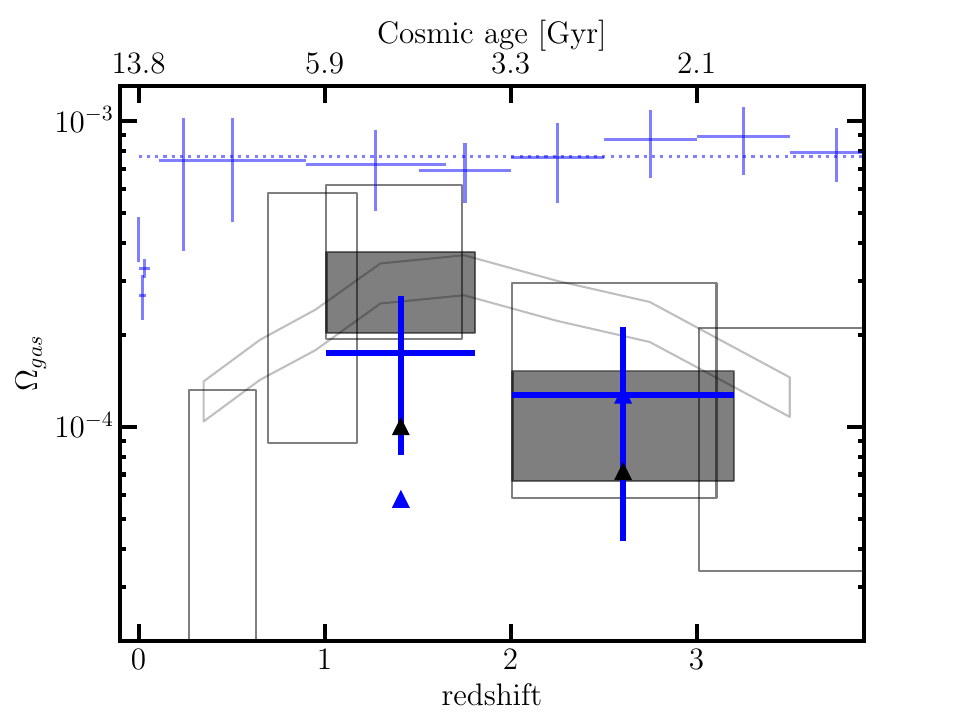}
    \includegraphics[width=0.49\textwidth]{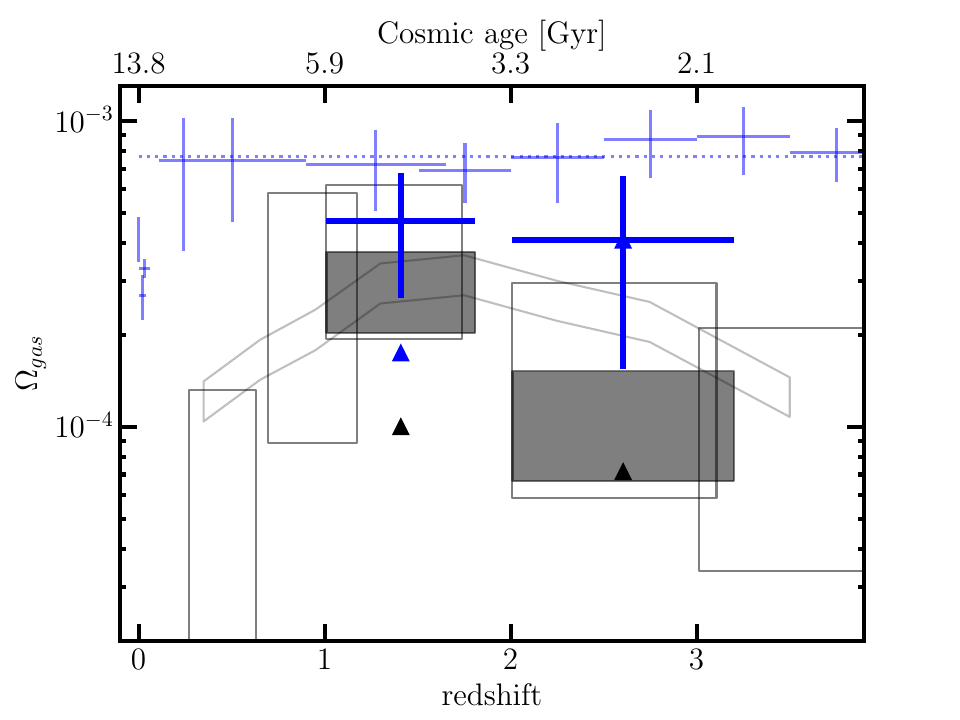}
    \caption{The ISM gas mass cosmological density evolution of its neutral and molecular phases. Left-hand side panel shows the results adopting the RT method, while the right-hand side panel shows those adopting the LR method. Blue error-bars refer to HI gas mass density measurements from the literature, while grey-line limited regions to H$_2$ gas mass literature estimates (see text for details). The thick line error-bars and grey filled boxes show our results. The blue and black triangles show the HI and H$_2$ gas mass densities, respectively, for sources with $L_{850}>10^{23}$\,W/Hz (LIRG-type galaxies).}
    \label{fig:mhievol}
\end{figure*}

The next step is thus to retrieve the cosmological mass density evolution of both dominant ISM gas constituents. In Figure~\ref{fig:mhievol} we show measurements from the literature and compare them with our results (left-hand side panel using the RT method, right panel using the LR one). In order to allow for a Cosmology-independent comparison with the literature, what is depicted in the figure is the cosmological mass density:
\begin{equation}
    \Omega_{gas} = \frac{\rho_{gas}}{\rho_{crit}}
\end{equation}
where
\begin{equation}
    \rho_{crit} = \frac{3~H_0^2}{8 \pi G}
\end{equation}
is today's Universe critical density. Blue color refers to HI gas, while black/gray refers to H$_2$. None of the data points have been corrected for heavier elements.

The assembly of literature measurements for ${\rm H_2}$ mass density are also of two types, either from CO detections \citep[][grey-line limited boxes]{Decarli20} or inferred from mm-continuum \citep[][continuous grey-line limited region]{Scoville17}. All data have been converted so that the same $\alpha_{\rm CO}$ is adopted (this also applies to the \citet{{Scoville17}} results which are based on a relation making use of a CO\,J:1-0 scaling and $\alpha_{\rm CO}=6.5$). Our results for H$_2$ mass density (gray-filled boxes) are quite in agreement with those from the literature (Appendix~\ref{app:overlap} shows the results if no ${\rm M_{H_2}}$-cut is applied between ASPECS and B21 samples). The fact that \citet{Scoville17} shows a higher cosmological content may be the result of sample incompleteness toward lower ${\rm M_{H_2}}$ affecting the CO-selected sample. The \citet{Scoville17} approach was to estimate ${\rm M_{\rm H_2}}$ with a dependence on redshift, stellar mass, and specific SFR. This allowed the team to account for a more representative population at cosmic noon \citep[see discussions in Sections 3.2 and 10 in][]{Scoville17}.

The thin blue crosses at the top of Figure~\ref{fig:mhievol} are the literature measurements for HI mass density \citep{Zwaan05,Lah07,Rao06,Martin10,Braun12,Zafar13}. We note that the measurements above redshift 0.3 are obtained via damped Lyman-$\alpha$ absorption systems, while the local Universe ($z\leq0.24$) measurements are direct detections of the 21\,cm line. Our results are shown as thick-line blue error-bars. Within the uncertainties, there is no significant evolution from $z\sim2.5$ and $z\sim1.5$. This happens irrespective of the HI-estimate method one adopts, the difference being that RT shows that CO/dust-selected samples recover $23\pm12\%$ and $17\pm11\%$ of the overall HI mass density at $z\sim1.5$ and $z\sim2.5$, respectively, while the LR method points to $61\pm27\%$ and $53\pm33\%$.

What both methods clearly show is that LIRG-type galaxies have a significant decrease in HI gas content during the same time range (by factors of $0.5\pm0.4$ and $0.4\pm0.3$ for RT and LR methods, respectively; data points shown as blue triangles). Doing the same exercise for the H$_2$ gas content for the same sub-sample (data points shown as black triangles), we find instead an increase of a factor of $1.4\pm0.8$, even though with less significance. This trend may explain the drop in SFR density seen for this kind of galaxies since $z\sim1$ \citep{LeFloch05,Goto11,Magnelli11}.

\iffalse
The difference remains even when one tries to recover as much as possible the mass density by fitting the MFs. Such estimates are shown by the rectangular dot-filled regions, and were obtained by integrating the fits to the MFs adopting a double power-law fit:
\begin{equation}
    ...
\end{equation}
The fits are shown in Appendix~\ref{app:mffits}. As one can see, some are not optimal, hence the reported regions should be just indicative, more so because at these redshifts, we are not complete to the overall galaxy population.
\fi

\subsection{Sources of contamination} \label{sec:contamin}

Actively accreting super-massive black holes (or Active Galactic Nuclei, AGN) may appear bright at radio wavelengths, to the point that their mm-emission is still dominated by synchrotron emission from the jet \citep[e.g.,][]{Messias21}. If that is the case, then the individual $L_{850}$ 
%${\rm M_{HI}}$ mm-
estimates will become overestimated, and those of ${\rm M_{HI}}$ will too as a result (Equations~\ref{eq:mhiRT} and \ref{eq:mhiLR}). This may result in a large impact in the cosmic mass density based on a small sample like the one used here.

To test this hypothesis we cross-matched the ASPECS and \citet{Birkin2021} samples with the radio catalogues reported by \citet[][UDS; 1.4\,GHz]{Simpson06}, \citet[][GOODS-South; 1.4\,GHz]{Kellermann08}, \citet[][COSMOS; 3\,GHz]{Smolcic17}. We only find matches with the \citet{Birkin2021} sample of sub-millimeter galaxies (SMGs): a total of 13 sources. One with a counterpart separation of 2\,arcsec, and the remainder at $<1\,$arcsec. Only 3 of the matched sources are used in Section~\ref{sec:mfdens}. Following the approach by \citet{Messias21}, we predict the contribution of the synchrotron emission at rest-frame 850\,$\mu$m adopting the radio flux estimates and a spectral index of $\alpha=-0.7$ (where $S_\nu\propto\nu^\alpha$). We find that 12 sources show an observed-to-predicted flux ratio of $\sim4-24$ (median of 13), while one (ALESS071.1) shows a ratio of 0.1, which is evidence for its mm emission to be synchrotron-dominated. %Note that these flux ratio estimates are generally lower-limits, since the adopted value for $L_{850}$ is estimated using photometry at observed 0.85--3\,mm wavelengths, with most detections at $<1.3\,$mm. Also, 
Curiously enough, the ${\rm M_{HI}}$ estimate for ALESS071.1 is actually negative (i.e., the ISM gas is mostly in its molecular phase) using either RT or LR methods. Finally, only three sources are also used in Section~\ref{sec:mfdens} (AS2COS0014.1, AS2UDS627.0, AS2UDS029.0), and for these we find ratios of 16, 9.8, 21 (respectively). Given the order-of-scale difference, we consider these sources not to have a significant synchrotron contribution to their luminosities at rest-frame 850\,$\mu$m, which also implies that our inferred HI cosmic mass densities are not contaminated (i.e., biased high) due to AGN contamination.

%% ==== DISCUSSION
\section{Discussion} \label{sec:disc}

\subsection{The evolution of the HI gas mass density}

\begin{figure}
    \centering
    \includegraphics[width=0.49\textwidth]{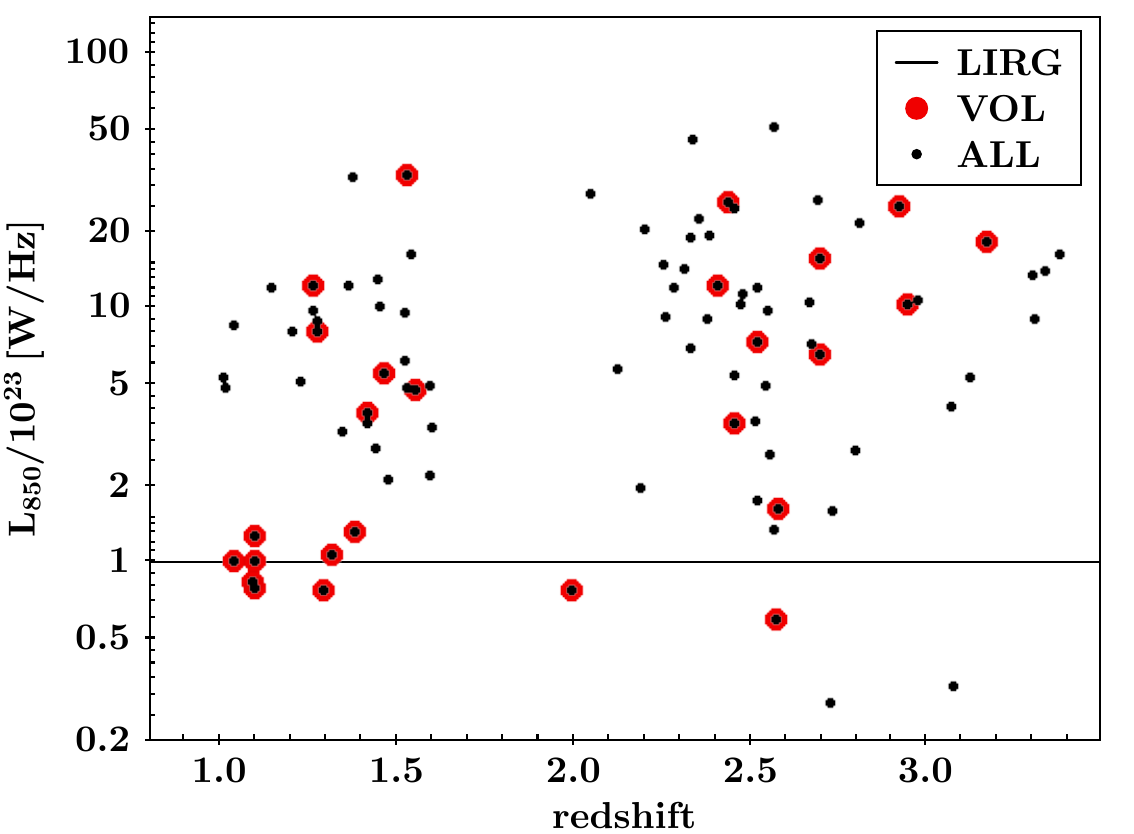}
    \caption{The distribution of $L_{850}$ with redshift for the sample considered in this work, zoomed-in to the redshift range of interest in Section~\ref{sec:mfdens}. The sample used for the cosmic gas density content ("VOL") is highlighted with red dots. For reference, the horizontal line shows the LIRG-like luminosity threshold.}
    \label{fig:ldustvsz}
\end{figure}

As mentioned in the previous section, the cosmological HI gas mass density shows no significant evolution for this CO/dust-selected sample at $1\lesssim z\lesssim3$. However, we have reasons to believe that this sample is significantly incomplete in the $z\sim2.5$ redshift range especially when we compare our H$_2$ content results with those from \citet{Scoville17} based on a purely dust-selected sample. If the missing population contributes significantly to the overall HI mass density, then the evolution may be a decreasing one with time from $z\sim2.5$ to $\sim$1.5. This is already apparent if one limits the analysis to the brightest galaxies: $L_{850}>10^{23}$\,W/Hz (for which our sample is expected to be somewhat complete, Figure~\ref{fig:ldustvsz}). This LIRG-like sub-sample shows a decrease in HI content of a factor of 2 during this time range, while an increase of a factor of 1.4 in H$_2$ gas content. This drop in neutral gas inflow may be the reason for the drop in SFR density seen for this kind of galaxies starting $\sim$3\,Gyr later, i.e., since $z\sim1$ \citep{LeFloch05,Goto11,Magnelli11}. The current uncertainties in our analysis are large enough to prevent us from determining how much HI is being converted into H$_2$ between these two redshift bins for this LIRG-like population (currently the errorbars are consistent with all missing HI being converted into H$_2$).

\subsection{ISM gas depletion times} \label{sec:deptimes}

In Figure~\ref{fig:gasratios}, we show that, overall, there is a 1-to-1 (RT) or 3-to-1 (LR) HI-to-${\rm H_2}$ content ratio, with no clear signs of evolution with redshift, except at the highest luminosities (Figure~\ref{fig:massfuncs}). This is thus a hint that the usual depletion times quoted in the literature taking into account only the ${\rm H_2}$ content are underestimated by a factor of 2 to 4. However, this is very much sample-selection dependent, since the CO/dust-selected sample used for the cosmological gas content evolution analysis shows median HI-to-${\rm H_2}$ content ratios of 0.9 and 0.5 at $z\sim1.5$ and $\sim$2.5, respectively, when using the RT method. Nevertheless, this still shows that a significant ISM gas content (at least a third) is not being considered for the estimate of gas depletion times.

For a quantitative example, we make use of the 12 SMGs reported in Section~\ref{sec:contamin} not to have significant AGN contamination in the radio ($1.2<z<4.8$). For these we estimated the SFR adopting the calibration based on the 1.4\,GHz continuum emission reported in \citet{Murphy11} \citep[that assumes the initial mass function from][]{Kroupa01}:
\begin{equation}
    {\rm SFR_{1.4\,GHz} [M_\odot~yr^{-1}] = 6.35\times10^{-22} ~ L_{1.4\,GHz} [W Hz^{-1}]}
\end{equation}

This retrieves a median SFR of 930\,M$_\odot$/yr (with a minimum--maximum of $\sim$500--2000\,M$_\odot$/yr). We find that the median depletion time ($\tau_{\rm SFR}={\rm M_{H_2}}$/SFR) considering only $\alpha_{\rm heavy}$M$_{\rm H_2}$ is $\sim410\,$Myr (with a minimum--maximum of $\sim$260--730\,Myr). If we instead consider the total ISM gas mass (Equation~\ref{eq:mism1}) adopting the RT method, the median $\tau_{\rm SFR}$ is now $\sim$880\,Myr ($\sim$350--1600\,Myr), while adopting the LR method, the values become $\tau_{\rm SFR}\sim2700\,$Myr ($\sim$710--5600\,Myr). These depletion time values assuming the total ISM gas mass are in line with the predictions from the cosmological hydrodynamic galaxy formation simulation of SMGs reported by \citet{Narayanan15}.

Moreover, in light of the latest findings on Giant Molecular Clouds (GMCs) lifetimes and SF efficiencies \citep[][and references therein]{Chevance23} we would like to raise awareness on the usage of the M$_{\rm H_2}$/SFR ratio to estimate $\tau_{\rm SFR}$ of a system. GMCs in the Milky-Way and other local galaxies are found to have lifetimes of $\tau_{\rm GMC}=10-30$\,Myr \citep{FeldmannGnedin11, Murray11, JeffresonKruijssen18}, their SF efficiency is found to be $\epsilon_{\rm SF}=2-10$\% \citep{Kruijssen19, Chevance20}, while the cloud mass distribution can be described with a power-law of the form dN/dM$\,\propto\,$M$^c$, where $-2.5<c<-1.5$ \citep{Murphy11,Mok20,Chevance23}.

We have ran a simple exercise to test how the GMC constraints impose a deviation from the simple M$_{\rm H_2}$/SFR ratio assumption. We first start randomly drawing GMC masses from the cloud mass distribution assuming $c=-2$ \citep[][and references therein]{Mok20} within a mass range of $10^4$ to $10^8$\,M$_\odot$ \citep[i.e., GMC scales;][]{Murray11,Chevance23}. In every iteration, we impose that the total cloud mass does not exceed the total mass in M$_{\rm H_2}$, and that the SFR at $\tau_{\rm GMC}$ lifetime scales (i.e., SFR per iteration: ${\rm SFR^{iter}_{GMC}= M^{iter}_{H_2}\times\epsilon_{SF}/\tau_{\rm GMC}}$) does not exceed the total SFR. The latter condition is to account for the fact that the commonly used long-wavelength or SED-based SFRs are values integrated over 100\,Myr \citep[Table 1 in][]{KennicuttEvans12}, while GMC lifetimes are 3 to 10 times shorter. So, that condition translates to: ${\rm SFR^{iter}_{GMC} \leq SFR_{tot} * \tau_{GMC}[Myr] / 100}$. After each iteration, we sum the adopted $\tau_{\rm GMC}$ to the cumulative of $\tau_{\rm SFR}$, and the used up molecular mass (M$^{\rm iter}_{\rm H_2}\times\epsilon_{\rm SF}$) is reduced from the ${\rm M_{H_2}^{tot}}$. The iterations continue until the leftover ${\rm M_{H_2}^{tot}}$ is $\leq10^6$\,M$_\odot$ (a scale at the level of the mass on a single GMC). This exercise shows that the M$_{\rm H_2}$/SFR ratio gives similar gas depletion timescales as adopting $\tau_{\rm GMC}=10$ or 20\,Myr and $\epsilon_{\rm SF}$ of 5 or 10\%, respectively. If one instead adopts the middle expected values for $\tau_{\rm GMC}$ (20\,Myr) and $\epsilon_{\rm SF}$ \citep[5\%; also the typical value in the Milky Way,][]{WilliamsMcKee97}, then one finds that $\tau_{\rm SFR}$ doubles with respect to the simple M$_{\rm H_2}$/SFR ratio. This assumption is roughly the same as when one randomly draws in each iteration values of $\tau_{\rm GMC}$ and $\epsilon_{\rm SF}$ assuming a flat probability distribution. As a result, we suggest the following statistical correction to $\tau_{\rm SFR}$ when estimated considering only M$_{\rm H_2}$:
\begin{equation}
    \tau_{\rm SFR} = C_1 * C_2 * \frac{\rm M_{H_2}}{{\rm SFR}}
\end{equation}
Where $C_1$ corrects for the existence of a HI reservoir, and $C_2$ corrects for the $\tau_{\rm GMC}$ and $\epsilon_{\rm SF}$ constraints. A value of $C_1=1$ is equivalent to a molecular gas dominated system, but the population wide results from this work point to $C_1\gtrsim2$ (Section~\ref{sec:gasfrac}). Note that these values of $C_1$ assume no extra time delay related to gas transitioning from a neutral to a molecular phase, nor molecular gas dissociation \citep[see section~3.2 in][]{Chevance23}. As for $C_2$, if we adopt average values for $\tau_{\rm GMC}$ and $\epsilon_{\rm SF}$, then $C_2=2$, but once again we note that no dependence was assumed for these two parameters on cloud mass or size \citep[see, for instance,][]{Mok20}. For that reason, we provide as supplementary material the code used to conduct this exercise in case the reader wishes to build upon the assumptions described above.

\subsection{Implications to Square Kilometre Array} \label{sec:ska}

\begin{figure}
    \centering
    \includegraphics[width=0.49\textwidth]{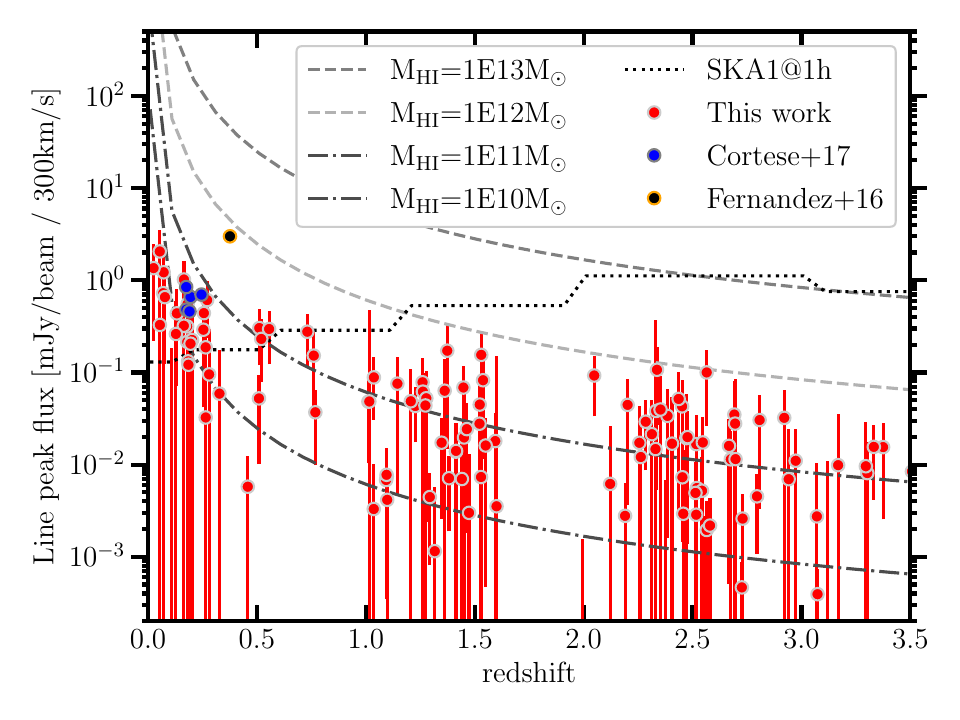}\\
    \includegraphics[width=0.49\textwidth]{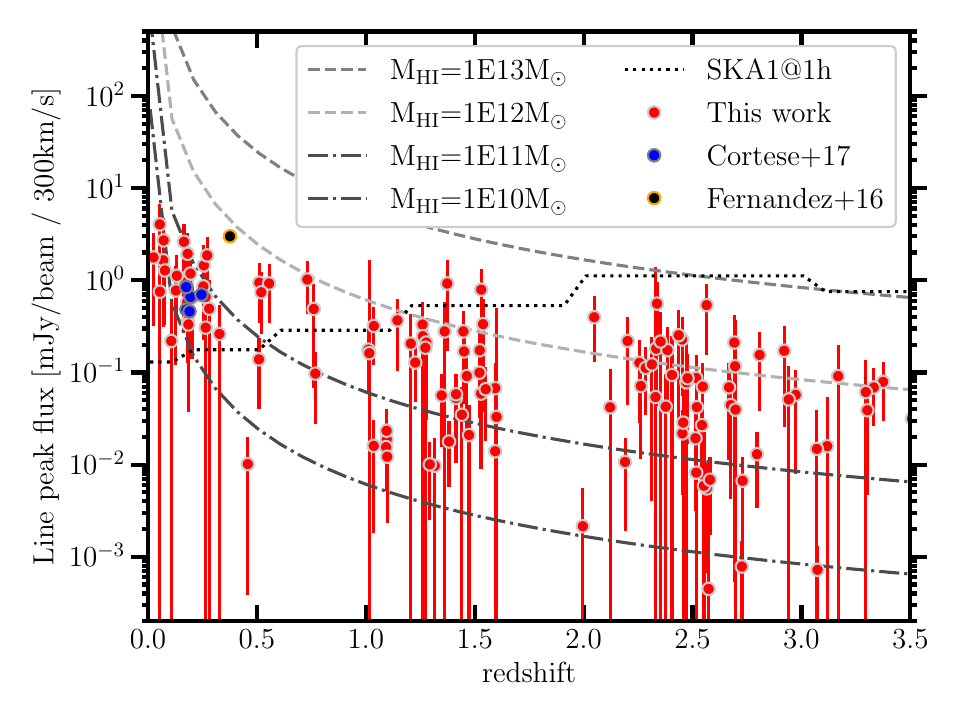}
    \caption{Predicted HI fluxes adopting the RT (top) and LR (bottom) methods for the sample considered in this work (red dots and errorbars). For reference, literature cases with direct HI detections at $0.2<z<0.5$ are also shown \citep{Cortese17,Fernandez16}, as well as expected fluxes for different M$_{\rm HI}$ contents with redshift ($\log(M_{\rm HI})=10$--13). The 1h on-source integration 3$\sigma$ sensitivity of SKA1 is shown as a dotted line.}
    \label{fig:ska}
\end{figure}

As a simple exercise, we make predictions of the HI flux for the sample we have gathered in this work. In order to do so, we use:
\begin{equation}
    {\rm S_{HI} = \frac{M_{HI} * (1+z)}{236 * D_l^2 * \Delta v}}
\end{equation}
where $\Delta$v is adopted to be 300~km/s and a Gaussian-like profile is assumed (even though a double peaked profile is more characteristic of typical galaxies). The line peak fluxes for both RT and LR methods are shown in Figure~\ref{fig:ska}. For reference, we also show the expected fluxes from different M$_{\rm HI}$ values with redshift. The expected SKA 1h on-source integration 3$\sigma$ sensitivity limit is shown \citep{Braun17,Braun19}. This figure shows that SKA will detect some of the galaxies assembled in this work at least up to $z\sim0.5$ within 1\,h. Above that redshift, it becomes more uncertain, with the both methods implying that, at $z>2$, SKA needs to spend at least $\sim100\,$h to directly detect the bulk of these CO- and continuum-selected galaxies. Based on the obtained HI mass functions (Section~\ref{sec:mfdens}, Figures~\ref{fig:massfuncs} and \ref{fig:massfuncsLR}), Table~\ref{tab:ska} resumes the expected number of galaxies directly detected by SKA, depending on time on source (ToS\,$=$\,1, 100\,h) and method used to estimate M$_{\rm HI}$ (RT and LR). We separate the estimates in redshift bins given the sensitivity differences in SKA Band~1 at 480--650\,MHz ($1.2<z<2.0$) and 350--480\,MHz ($2.0<z<3.1$)\footnote{SKA is expected to cover each of these frequency ranges in a single observation, so the ToS value is per frequency range reported.}. We do not present expected numbers for the $2.0<z<3.1$ case with ToS\,$=1$\,h, because the gas mass limit is regarded as unrealistic\footnote{For completeness, we refer that the results were nevertheless consistent with zero: $0^{+1.2}$ and $0.55^{+4.6}_{-0.42}$ for the RT and LR methods, respectively.} (see Section~\ref{sec:rtvslr}). Note that ToS\,$=100$\,h limits the direct detections to M$_{\rm HI}\gtrsim10^{11}\,$M$_\odot$, where we expect the analysis to be complete. Nevertheless, this sample is still limited by its CO- and continuum-detection nature, hence these numbers should be considered lower limits. Having this, large programs or stacking analysis will be required to characterize the star-forming population at these cosmic times.

\begin{table}
    \centering
    \begin{tabular}{cccccc}
    \hline
    \multicolumn{2}{c}{Scenario} & redshift & \multicolumn{2}{c}{Method} & Mass Limit \\
    ToS & $\Delta\nu$ & & RT & LR \\
    h & [MHz] & & \multicolumn{2}{c}{[deg$^{-2}$]} & [$\log(M_\odot)$] \\
    \hline \hline
    1 & 480--650 & 1.2--2.0 & $0.5^{+4.2}_{-0.38}$ & $3.0^{+5.4}_{-1.4}$ & 12 \vspace{0.1cm}\\
    %& 350--480 & 2.0--3.1 & $0^{+1.2}$ & $0.55^{+4.6}_{-0.42}$ & 13  \vspace{0.1cm}\\
    & 350--480 & 2.0--3.1 & --- & --- & 13  \vspace{0.1cm}\\
    100 & 480--650 & 1.2--2.0 & $120^{+140}_{-62}$ & $960^{+1200}_{-520}$ & 11  \vspace{0.1cm}\\
    & 350--480 & 2.0--3.1 & $7.9^{+31}_{-6.2}$ & $160^{+140}_{-72}$ & 12  \vspace{0.1cm}\\
    \end{tabular}
    \caption{Expected number of galaxies per deg$^2$ at $1<z<3$ directly detected by SKA based on the obtained HI mass functions (Section~\ref{sec:mfdens}, Figures~\ref{fig:massfuncs} and \ref{fig:massfuncsLR}). We show the numbers for different times on source (ToS) and adopting either RT or LR to estimate M$_{\rm HI}$. Note that 100\,h ToS limits the direct detections to M$_{\rm HI}\gtrsim10^{11}\,$M$_\odot$. We do not present expected numbers for the $2.0<z<3.1$ case with ToS\,$=1$\,h, because the gas mass limit is regarded as unrealistic (see Section~\ref{sec:rtvslr}).}
    \label{tab:ska}
\end{table}

\subsection{Reliability of RT and LR methods} \label{sec:rtvslr}

As shown in Figure~\ref{fig:dust2ism}, the dust-to-ISM gas calibration used local-Universe galaxies mostly with ISM gas masses below 10$^{11}\,$M$_\odot$, while some of the estimated M$_{\rm HI}$ values for the high-redshift sample reach one order of magnitude higher. Although baryonic masses reaching or in excess of 10$^{12}\,$M$_\odot$ may be expected for extreme cases \citep[see, for instance, the stellar mass functions at similar redshifts in][]{Weaver23}, we pursue an assessment of the credibility of each method. In this section, we put our results into perspective with respect to physical constrains that may potentially be regarded as strict.

For instance, the galaxy with the largest estimated HI mass content in the $1.0<z<3.2$ range (ALESS006.1 at $z_{\rm spec}=2.3368$) shows M$^{\rm RT}_{\rm HI}=(8.4\pm6.3)\times10^{11}\,$M$_\odot$ or M$^{\rm LR}_{\rm HI}=(4.4\pm3.2)\times10^{12}\,$M$_\odot$. The extreme M$_{\rm HI}$ value may be thought as associated with the low estimated continuum spectral-index ($\alpha=2.41$) resulting in an overestimate of the dust continuum, but the 3\,mm continuum detection actually directly traces the reference rest-frame wavelength 850\,$\mu$m. Also, from B21 and Section~\ref{sec:contamin}, there is no evidence for AGN contamination. Compared to its M$_{\rm H_2}=(3.2\pm1.3)\times10^{11}\,$M$_\odot$, and to the values reported by B21 for stellar and dust masses of ${\rm \log(M^*\,[M_\odot])=11.0\pm0.5}$ and ${\rm \log(M_{dust}\,[M_\odot])=9.31^{+0.04}_{-0.14}}$, respectively, we find HI fractions of 67$^{+20}_{-46}$\% and 91$^{+6}_{-30}$\% for the RT and LR methods, respectively. Although the uncertainties are large and significant larger HI fractions are expected with increasing redshift \citep{Chowdhury22}, the RT method estimate is regarded as more realistic, since the high HI fraction estimates by the LR method reach the expected properties of dwarf galaxies \citep{Huang12} or very high redshift samples \citep{Heintz22}, none of which are similar to ALESS006.1.

Another test that one can pursue is by making use of the tight relation between HI mass and the diameter of the HI disc \citep{BroeilsRhee97}. In \citet{Wang16} it is found to be ${\rm \log (D_{HI})=0.506\log(M_{HI}-3.293}$ with a scatter of 0.06\,dex. Directly applying this relation to the case of ALESS006.1 would imply a HI size of 550\,kpc (assuming M$^{\rm RT}_{\rm HI}$). Nevertheless, it is known that galaxies at high redshift are intrinsically smaller \citep{Trujillo07,Buitrago08,Buitrago23,vanderWel14,vanderWel24}. At the redshift of ALESS006.1, late-type galaxies are about 2.5 times smaller, while massive spheroids and/or quiescent galaxies can be factors of $4-7$ smaller. These values thus imply sizes for ALESS006.1 of $\sim$220\,kpc down to $\sim$80\,kpc. These values are still significantly larger than the observed stellar sizes at high redshifts \citep[e.g., $<20-30$\,kpc at $z\sim1$;][]{Buitrago23}), which is not surprising, but we must recall that the starting point of this work is a CO- and dust-based empirical relation, components which are expected to be comparable in size to the stellar component. However, we do note that if one considers that the molecular gas mass in ALESS006.1 (M$_{\rm H_2}=(3.2\pm1.3)\times10^{11}\,$M$_\odot$) was totally in its HI phase at a given point in the past, the minimum size assuming the above reasoning would still be $\sim$50\,kpc. At this point, we avoid addressing which relation does not hold at these redshifts --- the HI mass-size relation or the RT method --- but it is clear that the LR method implies much more extreme results, some deemed unrealistic.

Given the points presented in this section and results reported throughout the manuscript, we suggest the usage of the RT method over the LR one, and that the different sources of errors are taken into account (Section~\ref{sec:error}) in order to retrieve a realistic precision on the estimated HI mass based on the RT method.

%% ==== CONCLUSIONS
\section{Conclusions} \label{sec:conc}

In this work, we retrieve the atomic Hydrogen (HI) cosmological mass content at cosmic noon ($1.0<z<3.2$). We start by calibrating an empirical relation between millimeter continuum emission at rest-frame 850\,$\mu$m in galaxies and their inter-stellar medium (ISM) gas mass (${\rm M_{ISM}}$) with a local-Universe sample ($z<0.1$; see Section~\ref{sec:method}). Such relation has two flavors that we address throughout the manuscript: the so-called RT method assumes a simple ratio between the mm-continuum luminosity and ${\rm M_{ISM}}$, while the LR method assumes a log-log linear relation between the two. We then continue with applying this relation to a heterogeneous sample retrieved from the literature ($0.01<z<6.4$; Section~\ref{sec:sample}). With a special focus on a sub-set of galaxies selected in the millimeter wavelength range, namely being detected in CO\,(J$_{\rm up}=1,2,3$) and in continuum at 0.8--3\,mm (observed frame), we also derive the HI cosmological mass content at cosmic noon ($1.0<z<3.2$; Section~\ref{sec:mfdens}). Based on these results, we also present implications to the Square Kilometre Array (SKA). Overall, we conclude the following:

\begin{itemize}
    \item Based on the results and putting these into perspective with the literature, we give preference to the RT method over the LR (Section~\ref{sec:rtvslr});
    \item On a galaxy population wide view, we find no significant evolution on the atomic to molecular gas mass ratio, but specific sample selections may show differences with cosmic time (Section~\ref{sec:gasfrac});
    \item Based on this finding and in light of recent findings regarding the properties of Giant Molecular Clouds, we show that depletion times purely based in molecular gas content are underestimated by at least a factor of 2, but potentially by 4 (Section~\ref{sec:deptimes}).
    \item We find no significant evolution in the range $1.0<z<3.2$ of the HI cosmological gas content in (sub-)mm-selected galaxies, but this is a result of the sample selection being limited to more luminous/massive galaxies in the high-redshift range targeted here (Section~\ref{sec:mfdens});
    \item We find tentative evidence for a decrement in HI gas mass in luminous infrared galaxies (Section~\ref{sec:mfdens});
    \item Finally, we find that stacking analysis or large programs conducted with the SKA are required to study the bulk of the galaxy population referred in this manuscript where a 100\,h on-source project would directly detect $\sim$120 sources per square degree at $1.0<z<2.0$, and an order of scale less at $2.0<z<3.1$ (Section~\ref{sec:ska}).
\end{itemize}

\section*{Acknowledgements}
HM and JR acknowledge the ALMA-Princeton International Internship program, while HM and AG acknowledge the NRAO REU Chile Program and the SSDF ESO program, all of which allowed for the basis of this work to develop up to this stage.

HM would like to thank George Privon for the useful conversation about the recent results on GMRT HI surveys, and Kelley Hess for the useful feedback during the ALMA 10\,yr conference.

The team would like to thank the teams cited in Table~\ref{tab:sample-summary} for making their results and data public, without which this work would have not been possible.

This research made use of {\sc ipython} \citep{PerezGranger07}, {\sc numpy} \citep{Walt11}, {\sc matplotlib} \citep{Hunter07}, {\sc scipy} \citep{Virtanen20}, {\sc astropy} \citep[a community-developed core {\sc python} package for Astronomy,][]{Astropy13}, and {\sc topcat} \citep{Taylor05}.

This paper makes use of the following ALMA data: ADS/JAO.ALMA\#2016.1.00994.S, ADS/JAO.ALMA\#2017.1.01647.S, ADS/JAO.ALMA\#2017.1.00287.S. ALMA is a partnership of ESO (representing its member states), NSF (USA) and NINS (Japan), together with NRC (Canada), MOST and ASIAA (Taiwan), and KASI (Republic of Korea), in cooperation with the Republic of Chile. The Joint ALMA Observatory is operated by ESO, AUI/NRAO and NAOJ.

%%%%%%%%%%%%%%%%%%%%%%%%%%%%%%%%%%%%%%%%%%%%%%%%%%
\section*{Data Availability}
%The inclusion of a Data Availability Statement is a requirement for articles published in MNRAS. Data Availability Statements provide a standardised format for readers to understand the availability of data underlying the research results described in the article. The statement may refer to original data generated in the course of the study or to third-party data analysed in the article. The statement should describe and provide means of access, where possible, by linking to the data or providing the required accession numbers for the relevant databases or DOIs.

The information and data used in this manuscript is described in Section~\ref{sec:sample}. Specifically in Table~\ref{tab:sample-summary} we present the summary of the original works where the sample was retrieved from. Whenever needed, we also mention archival data used to complete the required information to pursue our work.



\bibliographystyle{mnras}
\bibliography{biblio} % if your bibtex file is called example.bib

%\begin{thebibliography}{99}
%\end{thebibliography}

%%%%%%%%%%%%%%%%%%%%%%%%%%%%%%%%%%%%%%%%%%%%%%%%%%

%%%%%%%%%%%%%%%%% APPENDICES %%%%%%%%%%%%%%%%%%%%%

\appendix

\section{Results with LR method} \label{app:lr}

\begin{figure}
    \centering
    \includegraphics[width=0.49\textwidth]{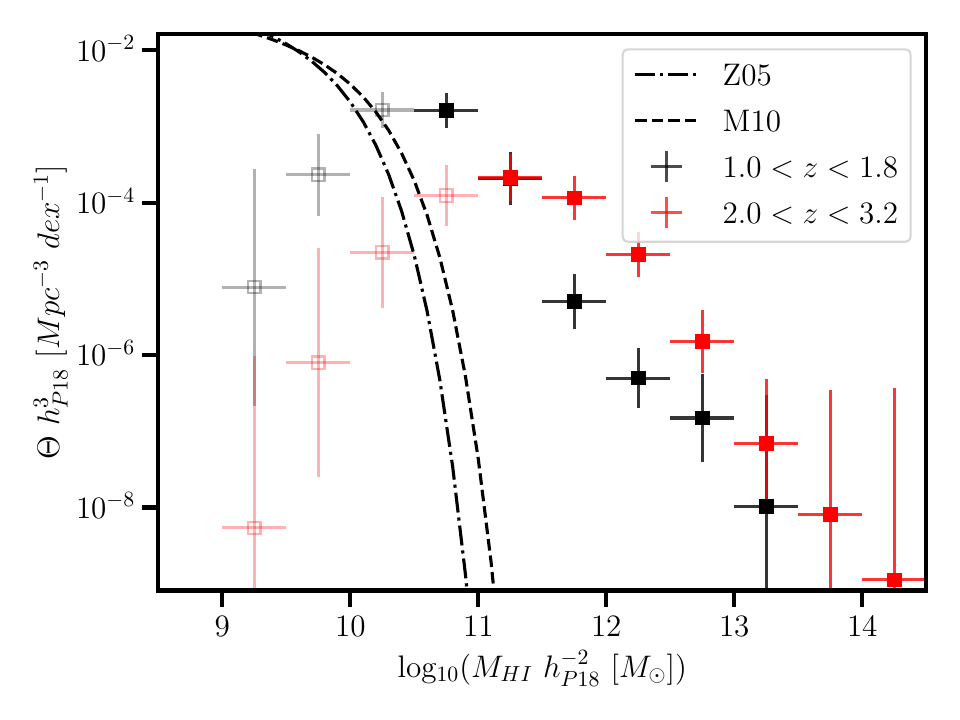}
    \caption{The same as in the top right-hand side panel of Figure~\ref{fig:massfuncs}, but adopting the LR method.}
    \label{fig:massfuncsLR}
\end{figure}

In this section, we complete Figure~\ref{fig:massfuncs} by showing the HI mass functions using the LR method in Figure~\ref{fig:massfuncsLR}.

\section{Results without the ${\rm M_{H_2}}$-cut} \label{app:overlap}

\begin{figure}
    \centering
    \includegraphics[width=0.49\textwidth]{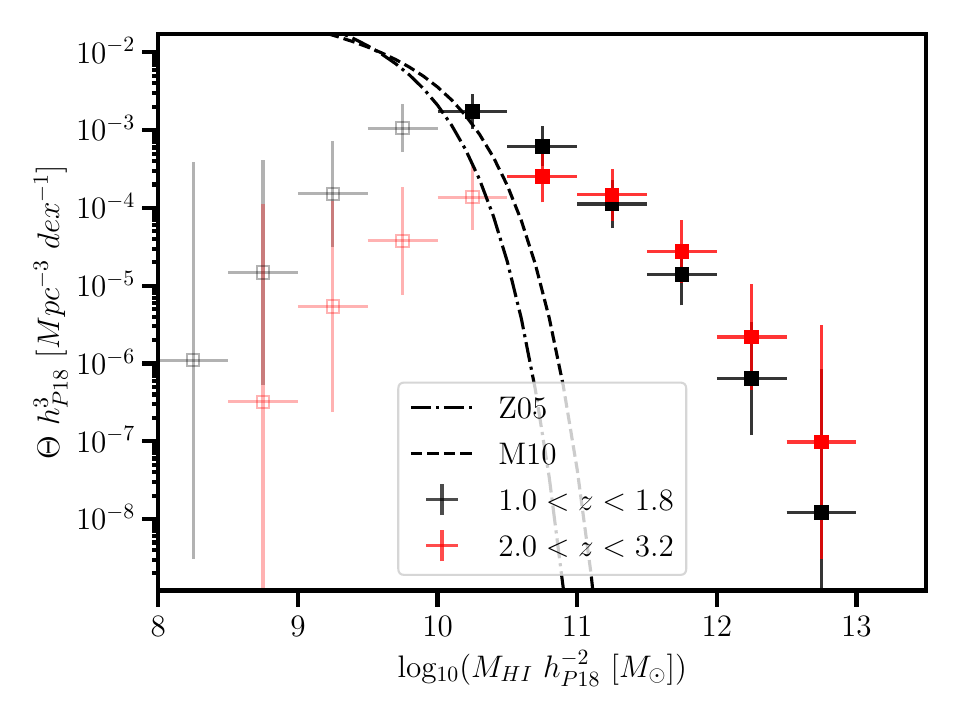}\\
    \includegraphics[width=0.49\textwidth]{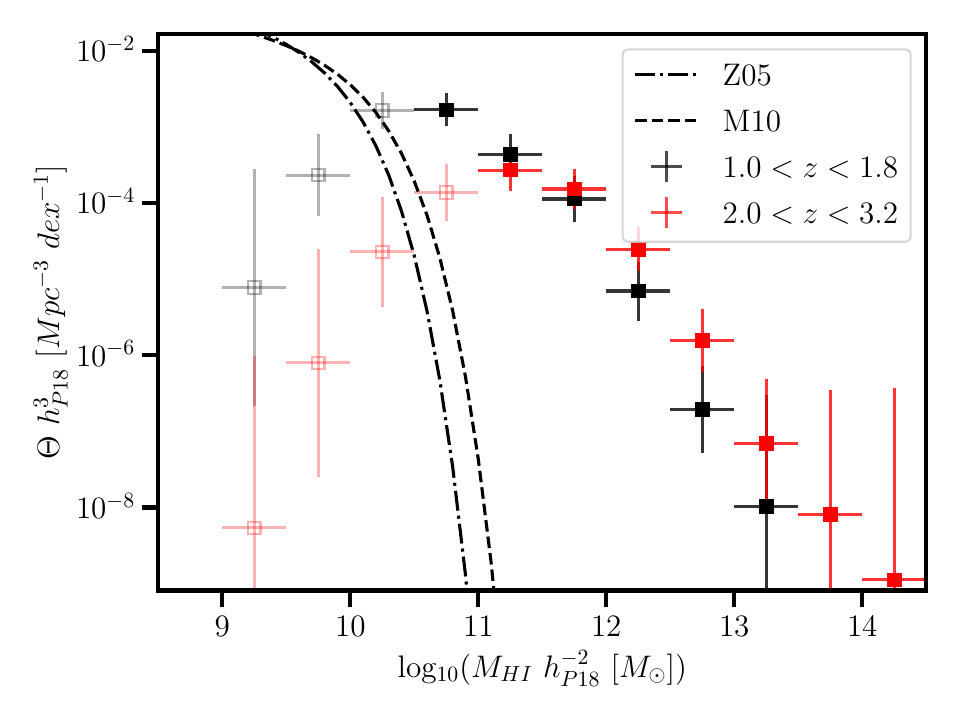}\\
    \caption{The same as in the top right-hand side panel of Figure~\ref{fig:massfuncs}, but not applying the cut in ${\rm M_{H_2}}$. Top panel considers the RT method, while the bottom one the LR method.}
    \label{fig:overlapRatio}
\end{figure}

\begin{figure}
    \centering
    \includegraphics[width=0.49\textwidth]{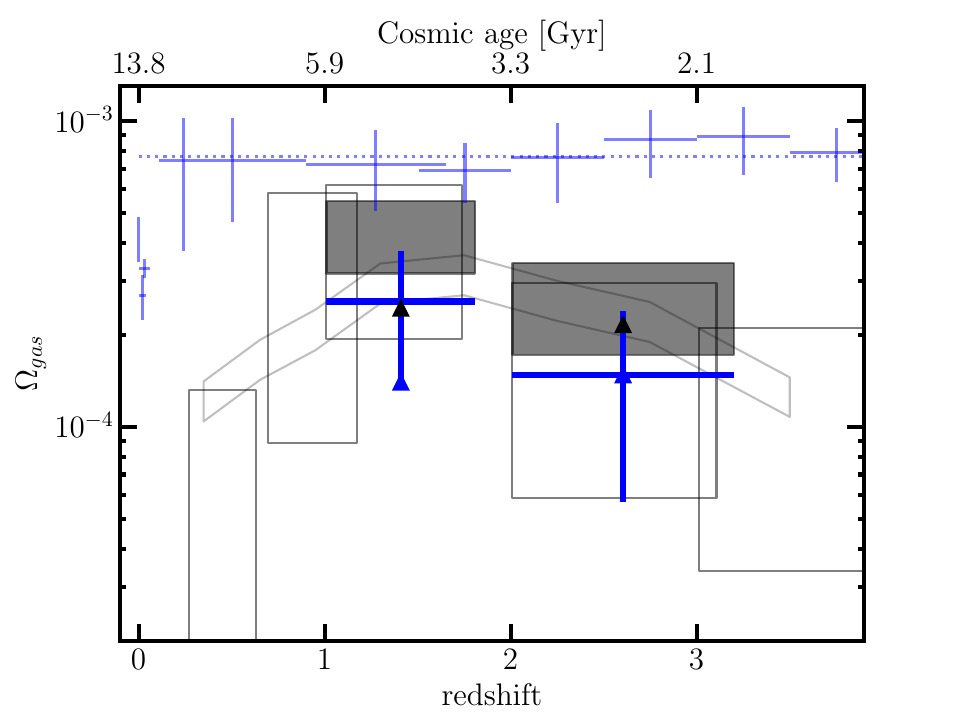}
    \includegraphics[width=0.49\textwidth]{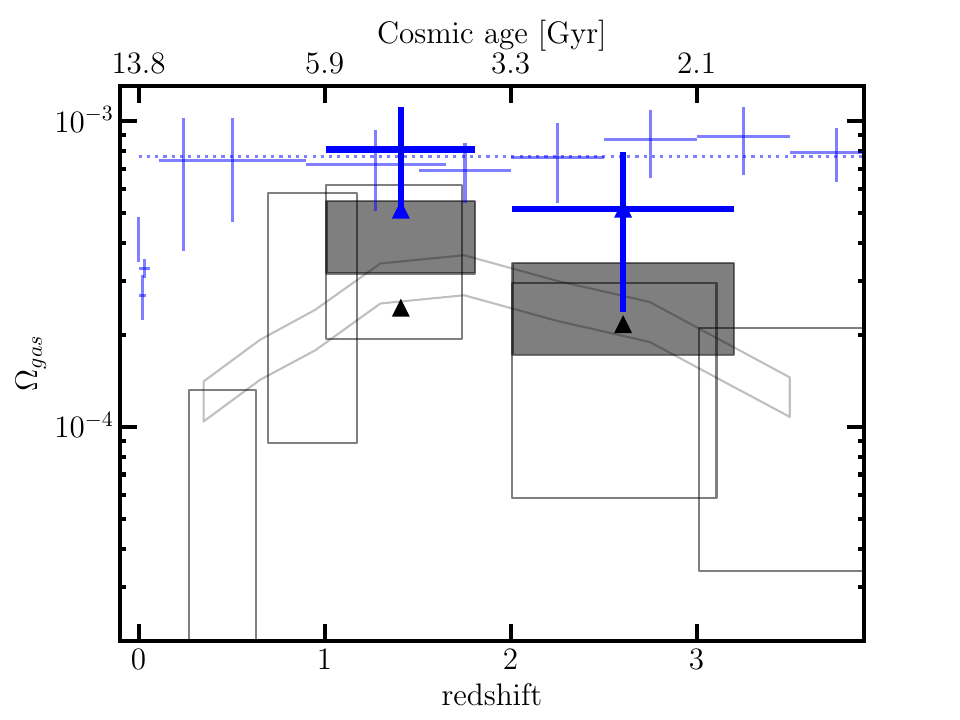}
    \caption{The same as in Figure~\ref{fig:mhievol}, but not applying the cut in ${\rm M_{H_2}}$. Top panel considers the RT method, while the bottom one the LR method.}
    \label{fig:overlapMHI}
\end{figure}

As mentioned in Section~\ref{sec:mfdens}, we have adopted a cut in ${\rm M_{H_2}}$ between samples ASPECS and B21 so to guarantee complementary between the two. Nevertheless, here we present the results (Figures~\ref{fig:overlapRatio} and \ref{fig:overlapMHI}) when such a cut is not applied. The main differences to highlight are: the expected increase in cosmic gas content in both gas phases (most noticeably in ${\rm M_{H_2}}$) and redshift bins; the LR method implies that, within the errors bars, the samples extracted from ASPECS and B21 already recover the ${\rm M_{HI}}$ content estimated from DLA systems analysis; the sample at $z\sim2.5$ still shows evidence for incompleteness with respect to the S17 approach; the ${\rm M_{HI}}$ content from the LIRG-like population is constant, but the ${\rm M_{H_2}}$ content still increases.

\section{Results without considering HI mass uncertainty} \label{app:nopdflf}

Given the large uncertainties associated to the HI mass estimate on an individual basis, we pursued a statistical analysis where the probability of a source to fall in a given mass bin was estimated adopting a log-normal probability distribution function (PDF) described by the most probably value and the associated error. In this section, we show the differences between this approach and the case if we would not consider the gas mass PDF, whereby a galaxy is associated to the bin comprising its most probable mass estimate value. In order to build such mass functions (MFs) with our reduced galaxy sample, we followed the standard strategy that works in the literature have been using to build CO luminosity functions \citep{Decarli19,Decarli20,Riechers19,Riechers20,Lenkic20}. Briefly, we use mass bins 0.5\,dex-wide separated by 0.2\,dex steps. Subsequent steps are not independent, but our goal here is to find significant deviations from the results when the mass PDF is considered.

\begin{figure*}
    \centering
    \includegraphics[width=0.49\textwidth]{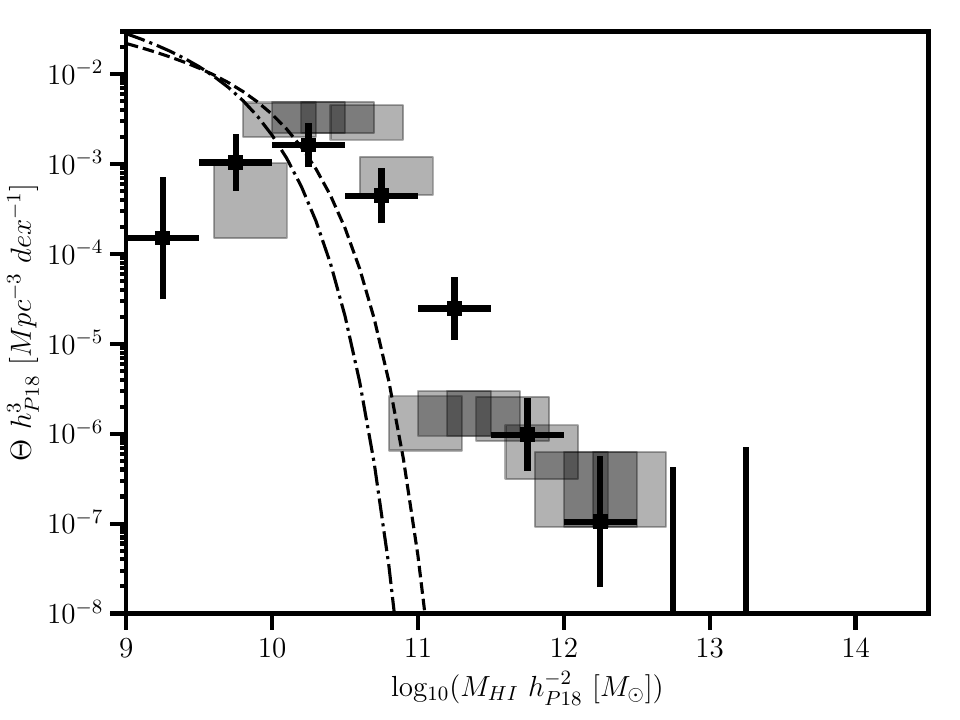}
    \includegraphics[width=0.49\textwidth]{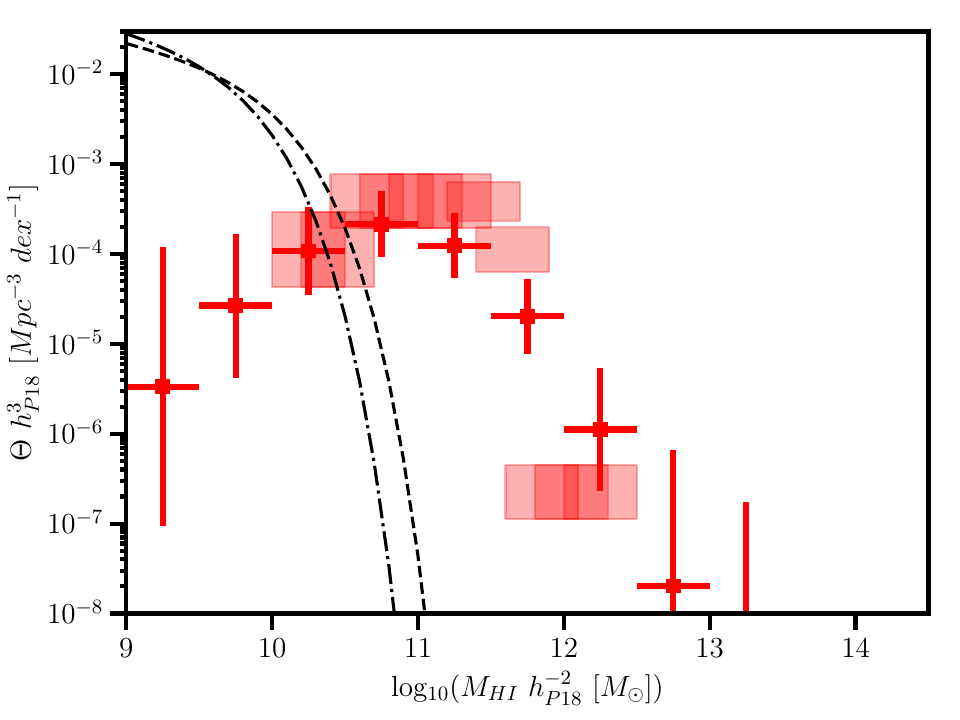}\\
    \includegraphics[width=0.49\textwidth]{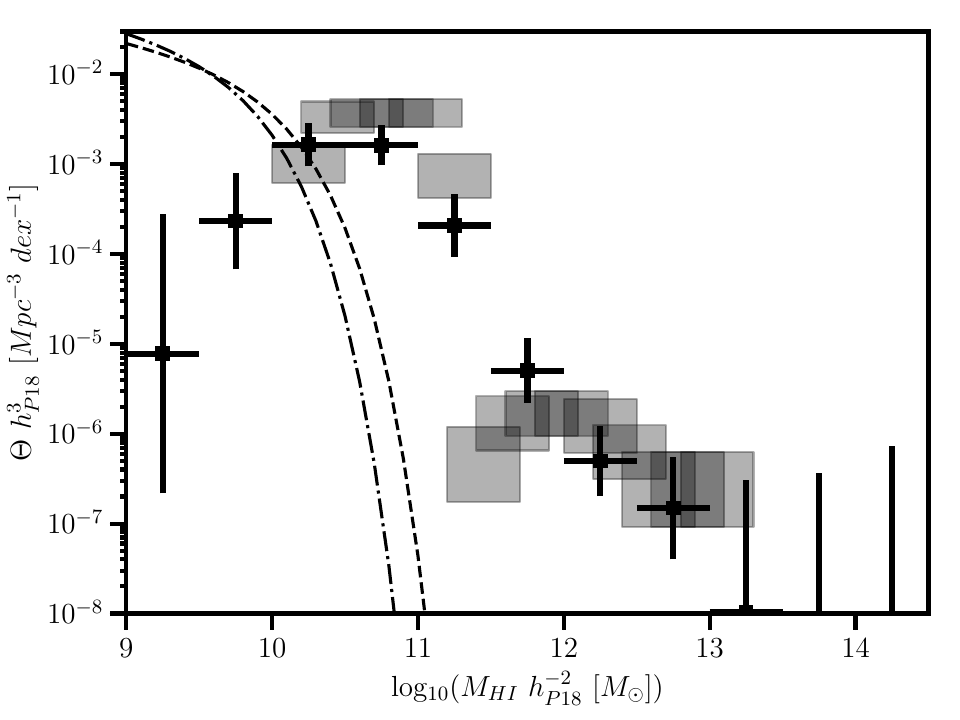}
    \includegraphics[width=0.49\textwidth]{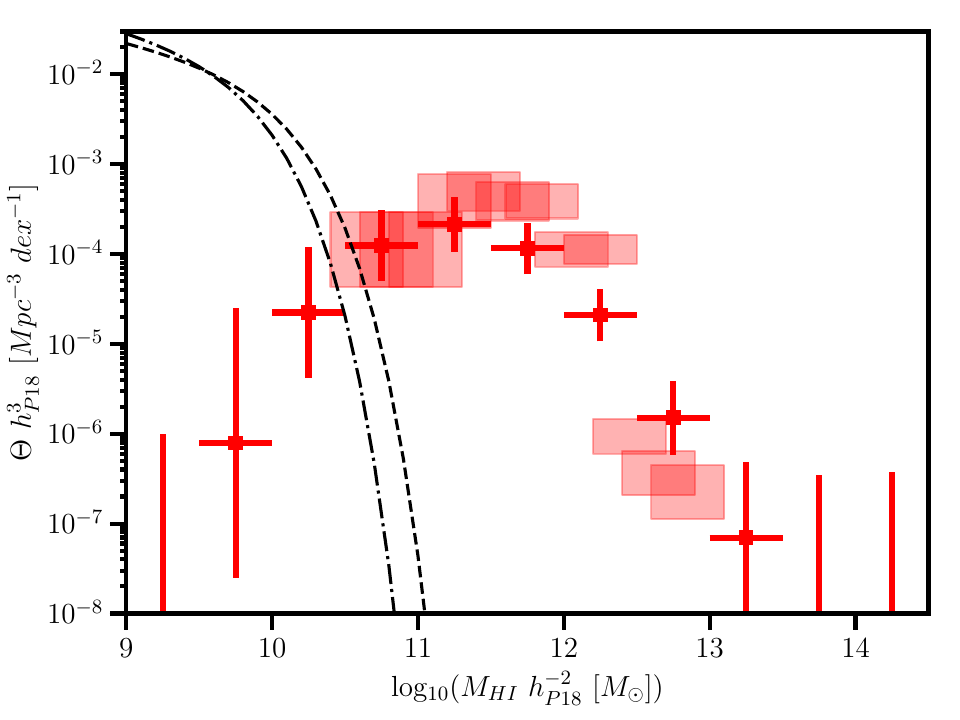}
    \caption{This figure compares the MFs when one distributes the source probability over different bins (Section~\ref{sec:mfdens}) \textit{versus} when one adopts only the highest probability value. The MFs adopting the latter approach were obtained by estimating the $\Sigma_i V^i_{max}$ in 0.5\,dex wide bins, at each step of 0.2\,dex \citep[following the usual approach in the literature to build CO LFs; e.g.,][]{Decarli19,Decarli20,Riechers19,Riechers20}. The top row shows the results adopting the RT method, while the bottom row those with LR. The left-hand side column shows the results for the $1.0<z<1.8$ range, while the right shows those for the $2.0<z<3.2$ range. The local HI mas functions are displayed for reference \citep{Zwaan05,Martin10}.}
    \label{fig:nopdflf}
\end{figure*}

In Figure~\ref{fig:nopdflf}, we directly compare both strategies at $1.0<z<1.8$ (left-hand side column) and $2.0<z<3.2$ (right column), and also for the RT (top row) and LR method (bottom row). For this exercise, we do not consider the cosmic variance error budget since we are comparing methods, while the sample is the same. As described in the main text, the expected trend resulting from using the gas mass PDF with respect to using single value is to a lower normalization and a flatter massive end. Although we see the tendency for the PDF analysis to be lower at the low-mass end, there is still agreement within the $1\sigma$ error bars, and we also see agreement at the massive end, especially in the lower-redshift range. Moreover, the light-end may be affected by lower-significance estimates that may be boosted by noise bias, and the PDF approach may be reducing this effect. However, we do not have the tools to confirm this point.

Based on this exercise, however, one can observe that the most extreme bins at the massive end do not trace ranges covered by the most probable values estimated to this sample. As a result, we intentionally remove them from the figures displayed in the main text. Specifically, we discard mass bins at ${\rm M_{HI}>10^{12.5}\,M_\odot}$ for the RT method, and at ${\rm M_{HI}>10^{13}\,M_\odot}$ for the LR method.

%%%%%%%%%%%%%%%%%%%%%%%%%%%%%%%%%%%%%%%%%%%%%%%%%%

% Don't change these lines
\bsp	% typesetting comment
\label{lastpage}
\end{document}